\documentclass[%
floatfix,
showkeys,
nofootinbib, %
superscriptaddress, %
]{revtex4-1}

\usepackage{cmap}

\usepackage{ucs}
\usepackage[utf8x]{inputenc}
\usepackage[T1,T2A]{fontenc}
\usepackage[german,russian,english]{babel}

\usepackage[sort&compress]{natbib}
\usepackage{amsmath}
\usepackage{amssymb}

\usepackage{mathtools}
\mathtoolsset{
showonlyrefs,
mathic = true
}

\allowdisplaybreaks

\usepackage{hyperref}
\hypersetup{backref,
 colorlinks=false}
\hypersetup{pdfborder=0 0 0}

\usepackage{breakurl}

\usepackage{microtype}
\UseMicrotypeSet[protrusion]{alltext}

\usepackage{listings}
\usepackage{listingsutf8}
\usepackage{fixltx2e}

\usepackage{nicefrac}

\makeatletter
\def\ps@pprintTitle{%
     \let\@oddhead\@empty
     \let\@evenhead\@empty
     \let\@oddfoot\@empty
     \let\@evenfoot\@oddfoot}
\makeatother

\usepackage{comment}
\usepackage{physics}
\usepackage{tensor}

\usepackage[section]{algorithm}
\usepackage{algorithmicx}
\usepackage{algpseudocode}
\usepackage{fancyvrb}

\usepackage{mathrsfs} 

\usepackage{amsthm} % Теоремы и т.п.
\usepackage{amsopn} % Пакет для объявления новых математических операторов \DeclareMathOperator

\newcommand{\const}{\mathrm{const}}

\input{listings-modelica.cfg}

\begin{document}
\graphicspath{{image/en/}}

\title{The Stochastic Processes Generation in OpenModelica}

\author{M. N. Gevorkyan}
\email{gevorkyan_mn@rudn.university}
\affiliation{Department of Applied Probability and Informatics,\\
  Peoples' Friendship University of Russia (RUDN University),\\
  6 Miklukho-Maklaya str., Moscow, 117198, Russia}

\author{A. V. Demidova}
\email{demidova_av@rudn.university}
\affiliation{Department of Applied Probability and Informatics,\\
  Peoples' Friendship University of Russia (RUDN University),\\
  6 Miklukho-Maklaya str., Moscow, 117198, Russia}

\author{A. V. Korolkova}
\email{korolkova_av@rudn.university}
\affiliation{Department of Applied Probability and Informatics,\\
  Peoples' Friendship University of Russia (RUDN University),\\
  6 Miklukho-Maklaya str., Moscow, 117198, Russia}

\author{D. S. Kulyabov}
\email{kulyabov_ds@rudn.university}
\affiliation{Department of Applied Probability and Informatics,\\
  Peoples' Friendship University of Russia (RUDN University),\\
  6 Miklukho-Maklaya str., Moscow, 117198, Russia}
\affiliation{Laboratory of Information Technologies\\
  Joint Institute for Nuclear Research\\
  6 Joliot-Curie, Dubna, Moscow region, 141980, Russia}

\author{L. A. Sevastianov}
\email{sevastianov_la@rudn.university}
\affiliation{Department of Applied Probability and Informatics,\\
  Peoples' Friendship University of Russia (RUDN University),\\
  6 Miklukho-Maklaya str., Moscow, 117198, Russia}
\affiliation{Bogoliubov Laboratory of Theoretical Physics\\
  Joint Institute for Nuclear Research\\
  6 Joliot-Curie, Dubna, Moscow region, 141980, Russia}

\begin{abstract}

\begin{description}
\item[Background] Component-based modeling language Modelica
  (OpenModelica is open source implementation) is used for the
  numerical simulation of complex processes of different nature
  represented by ODE system. However, in OpenModelica standard library
  there is no routines for pseudo-random numbers generation, which
  makes it impossible to use for stochastic modeling processes.
        
\item[Purpose] The goal of this article is a brief overview of a number
  of algorithms for generation a sequence of uniformly distributed
  pseudo random numbers and quality assessment of the sequence given
  by them, as well as the ways to implement some of these algorithms
  in OpenModelica system.

\item[Methods] All the algorithms are implemented in C language, and
  the results of their work tested using open source package
  DieHarder. For those algorithms that do not use bit operations, we
  describe there realisation using OpwnModelica. The other algorithms
  can be called in OpenModelica as C functions
\item[Results] We have implemented and tested about nine
  algorithms. DieHarder testing revealed the highest quality
  pseudo-random number generators. Also we have reviewed libraries
  Noise and AdvancedNoise, who claim to be adding to the Modelica
  Standard Library.
\item[Conclusions] In OpenModelica system can be implemented
  generators of uniformly distributed pseudo-random numbers, which is
  the first step towards to make OpenModelica suitable for simulation
  of stochastic processes.
\end{description}

\end{abstract}


\maketitle

\section{Introduction}

In this article we study the problem of generation uniformly
distributed pseudo-random numbers, stochastic Wiener and Poisson
processes in OpenModelica
framework~\cite{L_OpenModelica}. OpenModelica is one of the open
source implementation of Modelica~\cite{L_Modelica} modeling language
(for other implementations see~\cite{L_SciLab, L_LMS, L_Dymola,
  L_JModelica, L_WolframSystemModeler, L_MapleSim}). This language is
designed for modeling various systems and processes that can be
represented as a system of algebraic or differential equations. For
the numerical solution of the equations OpenModelica uses a number of
open source libraries~\cite{L_lis:2016, L_LAPACK:2016, L_UMFPACK:2016,
  L_KINSOL:2015}. However, in OpenModelica standard library there is
no any function even for generating uniformly distributed
pseudo-random numbers.

The first part of the article provides an overview of some algorithms
for generation pseudo-random numbers, including description of
pseudo-device \texttt{/dev/random} of Unix OS. For most of them we
provide the algorithm written in pseudocode. We implement all
described algorithms in the language of C and partly in
OpenModelica. Also we tested them with dieharder --- a random number
generator testing suite~\cite{L_DieHarder:2013}.

In the second part of the paper we describe algorithms for generating
normal and Poisson distributions. These algorithms are based on the
generators of uniformly distributed pseudo-random numbers. Then we
study the problem of computer generation of stochastic Wiener and
Poisson processes .

The third part of the article has a practical focus and is devoted to
the description of calling external functions, written in C language,
directly from OpenModelica programs code.

\section{Algorithms for generating uniformly distributed pseudo-random numbers}

In this section we will describe some of the most common generators of
uniformly distributed pseudo-random numbers. These generators are the
basis for obtaining a sequence of pseudo-random numbers of other
distributions.

\subsection{Linear congruential generator}

A linear congruential generator (LCG) was first proposed in 1949 by
D. H. Lehmer~\cite{L_DKnuth:1997:en}. The
algorithm~\ref{alg:lcg} is given by formula:
\[
        x_{n+1} = (a x_{n} + c) \mod m,\;\; n \geqslant 0,
\]
where $m$ is \emph{the mask} or \emph{the modulus} $m > 1$, $a$ is
\emph{the multiplier} $(0 \leqslant a < m)$, $c$ is \emph{the
  increment} $(0 \leqslant c < m)$, $x_0$ is \emph{the seed} or
initial value. The result of the repeated application of this
recurrence formula is \textit{linear congruential sequence}
$x_{1},\ldots,x_{n}$. A special case $c=0$ is called
\emph{multiplicative} congruential method.

\begin{algorithm}[H]
  \caption{\texttt{LCG} --- linear congruential generator}\label{alg:lcg}
  \begin{algorithmic}
    \Require $n$, $seed$
    \State $\const\;\;m \leftarrow 2^{64}$
    \State $\const\;\;a \leftarrow 6364136223846793005$
    \State $\const\;\;c \leftarrow 1442695040888963407$
    \State $x_{0} \leftarrow seed$
    \For{$i=0$ to $n$}
      \State $x_{i} = (a \cdot x_{i-1} + c) \mod m$
    \EndFor
    \State \Return $\{x_{0}, x_{1}, \ldots, x_{n}\}$
  \end{algorithmic}
\end{algorithm}

The number $m$, $a$, $c$ is called a <<magic>> or <<magic>> because
their values are specified in the code of the program and are selected
based on the experience of the use of the generator. The quality of
the generated sequence depends essentially on the correct choice of
these parameters. The sequence $\{x\}^{n}_{1}$ is periodic and its
period depends on the number $m$, which must therefore be large. In
practice, one chooses $m$ equal to the machine word size (for 32-bit
architecture --- $2^{32}$, for 64-bit architecture ---
$2^{64}$). D. Knuth~\cite{L_DKnuth:1997:en} recommends to choose
\[
a=6364136223846793005,\; c=1442695040888963407,\; m=2^{64} = 18446744073709551616.
\]
In the article~\cite{L_Ecuyer:1999}, you can find large tables with
optimal values $a$, $b$ and $m$.

Also there are generalisations of LCG, such as quadratic congruential
method $x_{n} = ( a x_{n-1}^2 + b x_{n-1} + d ) \mod m$ cubic
congruential method
$x_{n} = ( a x_{n-1}^3 + b x_{n-1}^2 + c x_{n-1} + d )\mod 2^e$.

Currently, the linear congruential method has mostly a historical
value, as it generates relatively low-quality pseudo-random sequence
compared to other, equally simple generators.

\subsection{Lagged Fibonacci generator}

The lagged Fibonacci generation can be considered as the
generalization of the linear congruential generator. The main idea of
this generalisation is to use multiple previous elements to generate
current one. Knuth~\cite{L_DKnuth:1997:en} claims that the first such
generator was proposed in the early 50-ies and based on the formula:
\[
        x_{n+1} = (x_{n} + x_{n-1}) \mod m.
\]
In practice, however, he showed himself not the best way. In 1958
George. J. Mitchell and D. Ph. Moore invented a much better
generator~\ref{alg:lfg}
\[
        x_{n} = (x_{n-n_{a}} + x_{n - n_{b}}) \mod m,\; n \geqslant \max(n_a, n_b).
\]
It was the generator that we now call \texttt{LFG} --- \textbf{l}agged
\textbf{F}ibonacci \textbf{G}enerator.

\begin{algorithm}[H]
  \caption{\texttt{LFG} --- Lagged Fibonacci generator}\label{alg:lfg}
  \begin{algorithmic}
    \State $n_{a} \leftarrow 55$
    \State $n_{b} \leftarrow 24$
    \Require $s_0, s_1,\ldots,s_{n_{b}}$, $n\geqslant  0$
    
    \State $x_{0},x_{1},\ldots,x_{n_{b}}$ $\leftarrow$ $r_0, r_1,\ldots,r_{n_{b}}$
    
    \For{$i = (n_{a}+1)$ to $n$}
      \If{$x_{i-n_a} \geqslant x_{i-n_b}$}
        \State $x_{i} = x_{i-n_a} - x_{i-n_b}$
      \ElsIf{$x_{i-n_a} < x_{i-n_b}$}
        \State $x_i = x_{i-n_a} - x_{i-n_b} + 1$
      \EndIf
    \EndFor
    \State \Return $\{x_{0},x_{1},\ldots,x_{n}\}$
  \end{algorithmic}
\end{algorithm}

As in the case of \texttt{LCG} generator <<magical numbers>> $n_a$ and
$n_b$ greatly affects the quality of the generated sequence. The
authors proposed to use the following magic numbers $n_a$ and $n_b$
\[
        n_a = 24, n_b = 55.
\]
Knuth~\cite{L_DKnuth:1997:en} gives a number of other values, starting
from $(37, 100)$ and finishing with $(9739, 23209)$. Period length of
this generator is exactly equal to $2^{e-1}(2^{55} - 1)$ when choosing
$m = 2^e$.

As can be seen from the algorithm for the initialization of this
generator must be used one an initial value and a sequence of
$\max(n_a, n_b)$ random numbers.

In open source GNU Scientific Library (GSL)~\cite{L_GSL:2015}
\emph{composite multi-recursive} generator are used. It was proposed
in paper~\cite{L_Ecuyer:1996}. This generator is a generalisation of
\texttt{LFG} may be expressed by the following formulas:
\[
        \begin{aligned}
                & x_{n} = (a_1 x_{n-1} + a_2 x_{n-2} + a_3 x_{n-3}) \mod m_{1},\\
                & y_{n} = (b_1 y_{n-1} + b_2 y_{n-2} + b_3 y_{n-3}) \mod m_{2},\\
                & z_{n} = (x_{n} - y_{n}) \mod m_{1}.
        \end{aligned}
\]
The composite nature of this algorithm allows to obtain a large period
equal to $10^{56} \approx 2^{185}$. The GSL uses the following
parameter values $a_{i}, b_{i}, m_{1}, m_{2}$:
\[
        \begin{array}{lll}
                a_1 = 0, & b_1 = 86098, & m_1 = 2^{32} - 1 = 2147483647,\\
                a_2 = 63308, & b_2 = 0, & m_2 = 2145483479,\\
                a_3 = -183326, & b_3 = -539608. &
        \end{array}
\]

Another method suggested in the paper~\cite{L_Ecuyer:1993:1} is also a
kind of Fibonacci generator and is determined by the formula:
\[
        x_{n} = (a_1 x_{n-1} + a_5 x_{n-5}) \mod 5,
\]
The GSL used the following values: $a_1 = 107374182$, $a_2 = 0$, $a_3 = 0$, $a_4 = 0$, $a_5 = 104480$, $m = 2^{31} - 1 = 2147483647$. The period of this generator is equal to $10^{46}$.
\subsection{Inversive congruential generator}
Inverse congruential method based on the use of inverse modulo of a number.
\[
        x_{i+1} = (a x^{-1}_{i} + b) \mod m
\]
where $a$ is \emph{multiplier} $(0\leqslant a < n)$, $b$ is
\emph{increment} $(0\leqslant b < n)$, $x_0$ is initial value
(seed). In addition $\text{GCD}(x_0,m)=1$ and $\text{HCF}(a,m)=1$ is
required.

This generator is superior to the usual linear method, however, is
more complicated algorithmically, since it is necessary to find the
inverse modulo integers which leads to performance reduction. To
compute the inverse of the number usually applies the extended
Euclidean algorithm~\cite[\S 4.3.2]{L_DKnuth:1997:en}.

\subsection{Generators with bitwise operations}
Most generators that produce high quality pseudo-random numbers
sequence use bitwise operations, such as conjunction, disjunction,
negation, exclusive disjunction (xor) and bitwise right/left shifting.

\subsubsection{Mersenne twister}

Mersenne twister considered one of the best pseudo-random
generators. It was developed in 1997 by Matsumoto and
Nishimura~\cite{L_Matsumoto:1998:MTE}. There are 32-,64-,128-bit
versions of the Mersenne twister. The name of the algorithm derives
from the use of Mersenne primes $2^{19937}-1$. Depending on the
implementation the period of this generator can be up to
$2^{216091}-1$.

The main disadvantage of the algorithm is the relative complexity and,
consequently, relatively slow performance. Otherwise, this generator
provides high-quality pseudo-random sequence. An important advantage
is the requirement of only one initiating number (seed). Mersenne
twister is used in many standard libraries, for example in the Python
3 module \texttt{random} ~\cite{L_Python3:3.5.1}.

Due to the complexity of the algorithm, we do not give its pseudocode
in this article, however, the standard implementation of the algorithm
created by Matsumoto and Nishimura freely available at the link
\url{http://www.math.sci.hiroshima-u.ac.jp/~m-mat/MT/emt64.html}.

\subsubsection{\texttt{XorShift} generator}
Some simple generators (algorithms~\ref{alg:xorshift64star}
and~\ref{alg:xorshift128plus}), giving a high quality pseudo-random
sequence were developed in 2003 by George. Marsala
(G. Marsaglia)~\cite{L_xorshift:2003, L_xorshift:2005}.

\begin{minipage}[t]{0.5\textwidth}
        \vspace{-10pt}
        %%%%%%%%%%%%%%%%%%%%%%%%%%%%%%%%%%%%%%%%%%%%%%%%%%%%%%%%%%%%%%%%%%%%%%%%%%%%%
        \begin{algorithm}[H]
                \begin{algorithmic}
                        \caption{\texttt{xorshift*}}\label{alg:xorshift64star}
                        \Require $n$, $seed$
                        \State $x \leftarrow seed$
                        \State $y_{0} \leftarrow x$
                        \For{$i=1$ to $n$}
                                \State $x$ $\leftarrow$ $x\; \oplus$ $x \gg 12$
                                \State $x$ $\leftarrow$ $x\; \oplus$ $x \ll 25$
                                \State $x$ $\leftarrow$ $x\; \oplus$ $x \gg 27$
                                \State $y_{i} \leftarrow  x \cdot 2685821657736338717$ 
                        \EndFor
                        \State \Return $\{y_{0}, y_{1}, \ldots, y_{n}\}$
                \end{algorithmic}
        \end{algorithm}
        %%%%%%%%%%%%%%%%%%%%%%%%%%%%%%%%%%%%%%%%%%%%%%%%%%%%%%%%%%%%%%%%%%%%%%%%%%%%%
\end{minipage}
\begin{minipage}[t]{0.5\textwidth}
        \vspace{-10pt}
        %%%%%%%%%%%%%%%%%%%%%%%%%%%%%%%%%%%%%%%%%%%%%%%%%%%%%%%%%%%%%%%%%%%%%%%%%%%%%
        \begin{algorithm}[H]
                \caption{\texttt{xorshift+}}\label{alg:xorshift128plus}
                \begin{algorithmic}
                        \Require $n$, $seed_{1}$, $seed_{2}$
                        \For{$i=1$ to $n$}
                                \State $x \leftarrow seed_{1}$
                                \State $y \leftarrow seed_{2}$
                                \State $seed_{1} \leftarrow y$
                                \State $x = x \oplus (x << 23)$
                                \State $seed_{2} = x \oplus y \oplus (x >> 17) \oplus (y >> 26)$
                                \State $z_{i} \leftarrow seed_{2} + y$
                        \EndFor
                        \State \Return $\{z_{1},\ldots,z_{n}\}$
                \end{algorithmic}
        \end{algorithm}
        %%%%%%%%%%%%%%%%%%%%%%%%%%%%%%%%%%%%%%%%%%%%%%%%%%%%%%%%%%%%%%%%%%%%%%%%%%%%%
\end{minipage}

\subsubsection{KISS generator}
Another group of generators (algorithms~\ref{alg:kiss}
and~\ref{alg:jkiss}), giving a high quality sequence of pseudo-random
numbers is KISS generators family~\cite{L_KISS:2011} (Keep It Simple
Stupid). They are used in the procedure \texttt{random\_number()} of
\texttt{Frotran} language (\texttt{gfortran}
compiler~\cite{L_gfortran:2015})

\begin{minipage}[t]{0.5\textwidth}
        \vspace{-10pt}
        %%%%%%%%%%%%%%%%%%%%%%%%%%%%%%%%%%%%%%%%%%%%%%%%%%%%%%%%%%%%%%%%%%%%%%%%%%%%%
        \begin{algorithm}[H]
                \begin{algorithmic}
                        \caption{\texttt{KISS}}\label{alg:kiss}
                        \Require $n$, $seed_{0},seed_{1},seed_{2},seed_{3}$
                        \State $t$
                        \For{$i=1$ to $n$}
                                \State $seed_{0} \leftarrow 69069 \cdot seed_{0} + 123456$
                                \State $seed_{1} \leftarrow seed_{1} \oplus (seed_{1} << 13)$
                                \State $seed_{1} \leftarrow seed_{1} \oplus (seed_{1} >> 17)$
                                \State $seed_{1} \leftarrow seed_{1} \oplus (seed_{1} << 5)$
                                \State $t \leftarrow 698769069 \cdot seed_{2} + seed_{3}$
                                \State $seed_{3} \leftarrow (t >> 32)$
                                \State $seed_{1} \leftarrow t$
                                \State $x_{i} \leftarrow seed_{0} + seed_{1} + seed_{2}$
                        \EndFor
                        \State \Return $\{x_{1},\ldots,x_{n}\}$
                \end{algorithmic}
        \end{algorithm}
        %%%%%%%%%%%%%%%%%%%%%%%%%%%%%%%%%%%%%%%%%%%%%%%%%%%%%%%%%%%%%%%%%%%%%%%%%%%%%
\end{minipage}
\begin{minipage}[t]{0.5\textwidth}
        \vspace{-10pt}
        %%%%%%%%%%%%%%%%%%%%%%%%%%%%%%%%%%%%%%%%%%%%%%%%%%%%%%%%%%%%%%%%%%%%%%%%%%%%%
        \begin{algorithm}[H]
                \begin{algorithmic}
                        \caption{\texttt{jKISS}}\label{alg:jkiss}
                        \Require $n$, $seed_{0},seed_{1},seed_{2},seed_{3}$
                        \State $t$
                        \For{$i=1$ to $n$}
                                \State $seed_{0} \leftarrow 314527869 \cdot seed_{0} + 1234567$
                                \State $seed_{1} \leftarrow seed_{1} \oplus (seed_{1} << 5)$
                                \State $seed_{1} \leftarrow seed_{1} \oplus (seed_{1} >> 7)$
                                \State $seed_{1} \leftarrow seed_{1} \oplus (seed_{1} << 22)$
                                \State $t \leftarrow 4294584393\cdot seed_{2} + seed_{3}$
                                \State $seed_{3} \leftarrow (t >> 32)$
                                \State $seed_{1} \leftarrow t$
                                \State $x_{i} \leftarrow seed_{0} + seed_{1} + seed_{2}$
                        \EndFor
                        \State \Return $\{x_{1},\ldots,x_{n}\}$
                \end{algorithmic}
        \end{algorithm}
        %%%%%%%%%%%%%%%%%%%%%%%%%%%%%%%%%%%%%%%%%%%%%%%%%%%%%%%%%%%%%%%%%%%%%%%%%%%%%
\end{minipage}

\subsection{Pseudo devices \texttt{/dev/random} and \texttt{/dev/urandom}}
\url{https://git.kernel.org/cgit/linux/kernel/git/stable/linux-stable.git/tree/drivers/char/random.c?id=refs/tags/v3.15.6#n52}

To create a truly random sequence of numbers using a computer, some
Unix systems (in particular GNU/Linux) uses the collection of
<<background noise>> from the operating system environment and
hardware. Source of this random noise are moments of time between
keystrokes (inter-keyboard timings), various system interrupts and
other events that meet two requirements: to be non-deterministic and
be difficult for access and for measurement by external observer.

Randomness from these sources are added to an "entropy pool", which is
mixed using a CRC-like function. When random bytes are requested by
the system call, they are retrieved from the entropy pool by taking
the SHA hash from it's content. Taking the hash allows not to show the
internal state of the pool. Thus the content restoration by hash
computing is considered to be an impossible task. Additionally, the
extraction procedure reduces the content pool size to prevent hash
calculation for the entire pool and to minimize the theoretical
possibility of determining its content.

External interface for the entropy pool is available as symbolic
pseudo-device \texttt{/dev/random}, as well as the system function:
\begin{lstlisting}[language=C]
  void get_random_bytes(void *buf, int nbytes);
  % \end{minted}
\end{lstlisting}
The device \texttt{/dev/random} can be used to obtain high-quality
random number sequences, however, it returns the number of bytes equal
to the size of the accumulated entropy pool, so if one needs an
unlimited number of random numbers, one should use a character
pseudo-device \texttt{/dev/urandom} which does not have this
restriction, but it also generates good pseudo-random numbers,
sufficient for the most non-cryptographic tasks.

\subsection{Algorithms testing}
A review of quality criterias of an sequence of pseudo-random numbers
can be found in the third chapter of the book~\cite{L_DKnuth:1997:en},
as well as in paper~\cite{L_Ecuyer:2007}. All the algorithms, which we
described in this articles, have been implemented in C-language and
tested with Dieharder test suite, available on the official
website~\cite{L_DieHarder:2013}.

\subsubsection{Dieharder overview}
Dieharder is tests suite, which is implemented as a command-line
utility that allows one to test a quality of sequence of uniformly
distributed pseudorandom numbers. Also Dieharder can use any generator
from GSL library~\cite{L_GSL:2015} to generate numbers or for direct
testing.
\begin{itemize}
        \item \verb|dieharder -l| --- show the list of available tests,
        \item \verb|dieharder -g -1| --- show the list of available random number generators; each generator has an ordinal number, which must be specified after \verb|-g| option to activate the desired generator.
        \begin{itemize}
                \item \verb|200 stdin_input_raw| --- to read from standard input binary stream,
                \item \verb|201 file_input_raw| --- to read the file in binary format,
                \item \verb|202 file_input| --- to read the file in text format,
                \item \verb|500 /dev/random| --- to use a pseudo-device \texttt{/dev/random},
                \item \verb|501 /dev/urandom| --- to use a pseudo-device \texttt{/dev/urandom}.
        \end{itemize}
\end{itemize}
Each pseudorandom number should be on a new line, also in the first lines of the file one must specify: type of number (\verb|d| --- integer double-precision), the number of integers in the file and the length of numbers ($32$ or $64$ - bit). An example of such a file:
\begin{verbatim}
                type: d
                count: 5
                numbit: 64
                1343742658553450546
                16329942027498366702
                3111285719358198731
                2966160837142136004
                17179712607770735227
\end{verbatim}
When such a file is created, you can pass it to \verb|dieharder|
\begin{verbatim}
dieharder -a -g 202 -f file.in > file.out
\end{verbatim}
where the flag \verb|-a| denotes all built-in tests, and the flag
\verb|-f| specifies the file for analysis. The test results will be
stored in \verb|file.out| file.

\subsubsection{Test results and conclusions}

\begin{center}
    \begin{tabular}{ | l | c | c | c |}
    \hline
    \textbf{The generator} & \textbf{Fail} & \textbf{Weak} & \textbf{Pass} \\ \hline\hline
                \texttt{LCG} & 52 & 6 & 55 \\ \hline
                \texttt{LCG2} & 51 & 8 & 54 \\ \hline
                \texttt{LFG} & 0 & 2 & 111 \\ \hline
                \texttt{ICG}  & 0 & 6 & 107 \\ \hline
                \texttt{KISS} & 0 & 3 & 110 \\ \hline
                \texttt{jKISS} & 0 & 4 & 109 \\ \hline
                \texttt{XorShift} & 0 & 4 & 109 \\ \hline
                \texttt{XorShift+} & 0 & 2 & 111  \\ \hline
                \texttt{XorShift*} & 0 & 2 & 111 \\ \hline
                \texttt{Mersenne Twister} & 0 & 2 & 111 \\ \hline
                \texttt{dev/urandom} & 0 & 2 & 111              \\ \hline
    \hline
    \end{tabular}
\end{center}

The best generators with bitwise operations are \texttt{xorshift*},
\texttt{xorshift+} and Mersenne Twister. They all give the sequence of
the same quality. The algorithm of the Mersenne Twister, however, is
far more cumbersome than \texttt{xorshift*} or \texttt{xorshift+},
thus, to generate large sequences is preferable to use
\texttt{xorshift*} or \texttt{xorshift+}.

Among the generators which use bitwise operations the best result
showed Lagged Fibonacci generator. The test gives results at the level
of \texttt{XorShift+} and Mersenne Twister. However, one has to set
minimum 55 initial values to initialize this generator, thus it's
usefulness is reduced to a minimum. Inverse congruential generator
shows slightly worse results, but requires only one number to initiate
the algorithm.

\section{Generation of Wiener and Poisson processes}

Let us consider the generation of normal and Poisson
distributions. The choice of these two distributions is motivated by
their key role in the theory of stochastic differential equations. The
most General form of these equations uses two random processes: Wiener
and Poisson~\cite{L_Platen_Bruti}. Wiener process allows to take into
account the implicit stochasticity of the simulated system, and the
Poisson process --- external influence.

\subsection{Generation of the uniformly distributed pseudo-random numbers from the unit interval}

Generators of pseudo-random uniformly distributed numbers are the
basis for other generators. However, most of the algorithms require a
random number from the unit interval $[0,1]$, while the vast majority
of generators of uniformly distributed pseudo-random numbers give a
sequence from the interval $[0, m]$ where the number $m$ depends on
the algorithm and the bitness of the operating system and processor.

To obtain the numbers from the unit interval one can proceed in two
ways. First, one can normalize existing pseudo-random sequence by
dividing each it's element on the maximum element. This approach is
guaranteed to give $1$ as a random number. However, this method is bad
when a sequence of pseudo-random numbers is too large to fit into
memory. In this case it is better to use the second method, namely, to
divide each of the generated number by $m$.

\subsection{Normal distribution generation}

An algorithm for normal distributed numbers generation has been
proposed in 1958 by George. Bux and
P. E. R. Mueller~\cite{L_BoxMuller:1958} and named in their honor
\textit{Box-Muller transformation}. The method is based on a simple
transformation. This transformation is usually written in two formats:
\begin{itemize}
  \item standard form (was introduce in the paper~\cite{L_BoxMuller:1958}),
  \item polar form (suggested by George Bell~\cite{L_Bell:1968} and R. Knop~\cite{L_Knop:1969}).
\end{itemize}

\textbf{Standard form.} Let $x$ and $y$ are two independent, uniformly
distributed pseudo-random numbers from the interval $(0, 1)$, then
numbers $z_1$ and $z_2$ are calculated according to the formula
\[
  z_{1} = \cos(2\pi y)\sqrt{-2\ln{x}},\; z_{2} = \sin(2\pi y)\sqrt{-2\ln{x}}
\]
и are independent pseudo-random numbers distributed according to a
standard normal law $\mathcal{N}(0,1)$ with expectation $\mu = 0$ and
the standard deviation $\sigma = 1$.

\textbf{Polar form.} Let $x$ and $y$ --- two independent, uniformly
distributed pseudo-random numbers from the interval $[-1, 1]$. Let us
compute additional value $s = x^2 + y^2$. If $s>1$ or $s=0$ then
existing $x$ and $y$ values should be rejected and the next pair
should be generated and checked. If $0 < s \geqslant 1$ then the
numbers $z_1$ and $z_2$ are calculated according to the formula
\[
  z_1 = x\sqrt{\dfrac{-2\ln{s}}{s}},\; z_2 = y \sqrt{\dfrac{-2\ln{s}}{s}}
\]
and are independent random numbers distributed according to a standard
normal law $\mathcal{N}(0,1)$.

For computer implementation is preferable to use a polar form, because
in this case one has to calculate only single transcendental function
$\ln$, while in standard case three transcendental functions ($\ln$,
$\sin$ $\cos$) have to be calculated. An example of the algorithm
shown in figure~\ref{fig:normal}

To obtain a general normal distribution from the standard normal
distribution, one can use the formula $Z = \sigma \cdot z + \mu$ where
$z \sim \mathcal{N}(0,1)$, and $Z \sim \mathcal{N}(\mu,\sigma)$.

  % y = - \dfrac{1}{\lambda}\ln(1-x),

\subsection{The generation of a Poisson distribution}
To generate a Poisson distribution there is a wide variety of
algorithms~\cite{L_Devroye, L_Ahrens:1974, L_Ahrens:1982}. The easiest
was proposed by Knut~\cite{L_DKnuth:1997:en}. This
algorithm~\ref{alg:poisson} uses uniform pseudo-random number from the
interval $[0,1]$ for it's work. The algorithm's output example is
depicted on figure~\ref{fig:poisson}
\begin{algorithm}[H]
  \caption{The generator of the Poisson distribution}\label{alg:poisson}
  \begin{algorithmic}
                \Require $seed$, $\lambda$
    
                \State $\Lambda \leftarrow \exp(-\lambda)$, $k \leftarrow 0$, $p \leftarrow 1$, $u \leftarrow seed$
    \Repeat
      \State $k \leftarrow k + 1$
                        \State $u \leftarrow rand(u)$ \Comment{generation of uniformly distributed random number}
                        \State $p = p \cdot u$
    \Until{$p > \Lambda$}
    \State \Return $k - 1$
  \end{algorithmic}
\end{algorithm}

\begin{figure}
  \centering
  \begin{minipage}[b]{0.48\linewidth}
    \includegraphics[width=1.0\linewidth]{normal.pdf}
    \caption{Normal distribution}
    \label{fig:normal}
  \end{minipage}
  \hfill
  \begin{minipage}[b]{0.48\linewidth}
    \includegraphics[width=1.0\linewidth]{poisson.pdf}
    \caption{Poisson distribution}
    \label{fig:poisson}
  \end{minipage}
\end{figure}

\subsection{Generation of Poisson and Wiener processes}

Now we going to use generators of normal and Poisson distributions to
generate Wiener and Poisson stochastic processes. Let us first give
the definition of these processes and then proceed to algorithms
descriptions.

\subsubsection{Definition of Wiener and Poisson stochastic processes}

Let $(\Omega, \mathscr{A}, P)$ --- probability space, where $\Omega$
--- the space of elementary events $\mathscr{A}$ --- $\sigma$ -
algebra of subsets of $\Omega$ (a random event), $P$ --- the
probability or probability measure, such that $P(\Omega) = 1$.

A family of random variables
$X = \{X_{t}, 0 \leqslant t \leqslant T\}$, where
$X_{t} \in \mathbb{R}^{d}$ will be called $d$--dimensional stochastic
process, where a set of finite-dimensional distribution functions have
the following properties (see~\cite{L_Kloeden_Platen, L_Oksendal_en})
        \[
                F_{X_{t_{1}},X_{t_{2}},\ldots,X_{t_{k}}}(x_{i_1},x_{i_2},\ldots,x_{i_k}) = P(X_{t_{1}}\leqslant x_{i_1}, X_{t_{2}} \leqslant x_{i_2}, \ldots, X_{t_{k}} \leqslant x_{i_k})
        \]
        for all $i_{k} = 1,2,3,\ldots$, $k = 1,2,3,\ldots$, $x_{i} \in \mathbb{R}^{d}$ и $t_{k} \in \mathrm{T}$

        The state space of $X$ is called $d$--dimensional Euclidean
        space $\mathbb{R}^{d}$, $d=1,2,3,\ldots$. The time interval
        $[0, T]$, where $T>0$. In numerical methods the sequence of
        time moments $\{t_0,t_1,t_2,\ldots \}$ is used.

        Random piecewise-constant process
        $N = \{N_{t}, 0 \leqslant t \leqslant T\}$ with intensity
        $\lambda > 0$ is called the \emph{Poisson process} if the
        following properties are true (see~\cite{L_Kloeden_Platen,
          L_Oksendal_en}):
        \begin{enumerate}
                \item $P\{N_0 = 0\} = 1$, otherwise $N_0 = 0$ almost surely.
                \item $N_t$ has independent increments: $\{\Delta N_{0}, \Delta N_{1}, \ldots\}$ are independent random variables; $\Delta N_{t_i} = N_{t_{i+1}} - N_{t_{i}}$ and $0 \leqslant t_{0} < t_{1} < t_{2} < \ldots  < t_{n} \leqslant T$; $\Delta N_{t_i} = N_{t_{i+1}} - N_{t_{i}}$ и $0 \leqslant t_{0} < t_{1} < t_{2} < \ldots  < t_{n} \leqslant T$.
                \item There is a number $\lambda > 0$ such as, for any increment $\Delta N_{i},\; i = 0, \ldots, n-1$, $ E[\Delta N_{i}]= \lambda \Delta t_{i}$.
                \item If $P(s) = P\{N_{t+s} - N_{t} > 2\}$, then $\lim\limits_{s\to 0} \dfrac{P(s)}{s} = 0$.
        \end{enumerate}

        The random process $W = \{W_t, 0 \leqslant t \leqslant T\}$ is called scalar \textit{Wiener process} (Wiener) if the following conditions are true (see~\cite{L_Kloeden_Platen, L_Oksendal_en}).
        \begin{enumerate}
                \item $P\{W_0 = 0\} = 1$, otherwise $W_0 = 0$ almost surely.
                \item $W_t$ has independent increments: $\{\Delta W_{0}, \Delta W_{1}, \ldots\}$ are independent random variables; $\Delta W_{t_i} = W_{t_{i+1}} - W_{t_{i}}$ и $0 \leqslant t_{0} < t_{1} < t_{2} < \ldots  < t_{n} \leqslant T$.
                \item $\Delta W_{i} = W_{t_{i+1}} - W_{t_{i}} \sim \mathcal{N}(0,t_{i+1}-t_{i})$ где $0\leqslant t_{i+1} < t_{i} < T$, $i=0,1,\ldots,n-1$.
        \end{enumerate}

        From the definition it follows that $\Delta W_{i}$ is normally
        distributed random variable with expectation
        $\mathbb{E}[\Delta W_{i}] = \mu = 0$ and variance
        $\mathbb{D}[\Delta W_{i}] = \sigma^{2} = \Delta t_{i}$.

        Wiener process is a model of \emph{Brownian motion} (the
        random walk). The process $W_t$ in following time points
        $0 = t_{0} < t_{1} < t_{2} < \ldots < t_{N-1} < t_{N}$
        experiences random additive changes:
        $ W_{t_{1}} = W_{t_{0}} + \Delta W_0,\; W_{t_{2}} = W_{t_{1}}
        + \Delta W_1,\; \ldots,\; W(t_{N}) = W(t_{N-1}) + \Delta
        W_{n-1}, $ where
        $\Delta W_{i} \sim \mathcal{N}(0,\Delta t_{i})$,
        $\forall i = 0,\ldots,n-1$.

Let us write $W_{t_{n}}$ in the following form:
$
        W_{t_{n}} = W_{t_{0}} + \sum\limits_{i=0}^{n-1}\Delta W_{i} 
$
and consider that $\mathbb{E}[\Delta W_{i}] = 0$ and $\mathbb{D}[\Delta W_{i}] = \Delta t_{i}$. We can show now, that the sum of normally distributed random numbers $\Delta W_{i}$ is also normally distributed random number:
\[
        \mathbb{E}\sum\limits_{i=0}^{n-1}\Delta W_{i} = 0, \;\; \mathbb{D}\sum\limits_{i=0}^{n-1}\Delta W_{i} = \sum\limits_{i=0}^{n-1} \Delta t_{i} = T,\;\; \sum\limits_{i=0}^{n-1}\Delta W_{i} \sim \mathcal{N}(0,T).
\]

Multidimensional Wiener process
$\mathbf{W}_{t}\colon \Omega\times[t_{0},T]\to \mathbb{R}^{m}$ is
defined as a random process composed of jointly independent
one-dimensional Wiener processes $W^{1}_t,\ldots,W^{m}_t$. Increments
$\Delta W^{\alpha}_{i},\;\forall \alpha = 1,\ldots,m$ are jointly
independent normal distributed random variables. On the other hand,
the vector $\Delta W^{\alpha}_{i}$ can be represented as a
multidimensional normally distributed random variable with expectation
vector $\mu = 1$ and a diagonal covariance matrix.

\subsubsection{The generation of the Wiener process}

To simulate one-dimensional Wiener process, one should generate the
$N$ normally distributed random numbers
$\varepsilon_{1}, \ldots, \varepsilon_{N}$ and build their cumulative
sums of $\varepsilon_{1}$, $\varepsilon_{1} + \varepsilon_{2}$,
$\varepsilon_{1} + \varepsilon_{2} + \varepsilon_{3}$. As result we
will get \emph{a trajectory} of the Wiener process
$W(t)$~cm~Fig.~\ref{fig:wienerprocess}.

In the case of multivariate random process, one needs to generate $m$ sequences of $N$ normally distributed random variables.

\subsubsection{The generation of a Poisson process}

A simulation of the Poisson process is much like Wiener one, but now
we need to generate a sequence of numbers distributed according to the
Poisson law and then calculate their cumulative sum. The plot of
Poisson process is shown in Fig.~\ref{fig:poissonprocess}. The figure
shows that the Poisson process represents an abrupt change in numbers
have occurred over time events. The intensity $\lambda$ depends on the
average number of events over a period of time.

Because of this characteristic behavior of the Poisson process is also
called an process with jumps and stochastic differential equations,
with Poisson process as second driving process, are called equations
with jumps~\cite{L_Platen_Bruti}

\begin{figure}
  \centering
  \begin{minipage}[b]{0.48\linewidth}
    \includegraphics[width=1.0\linewidth]{wiener_process.pdf}
    \caption{Wiener process}
    \label{fig:wienerprocess}
  \end{minipage}
  \hfill
  \begin{minipage}[b]{0.48\linewidth}
    \includegraphics[width=1.0\linewidth]{poisson_process.pdf}
    \caption{Poisson process}
    \label{fig:poissonprocess}
  \end{minipage}
\end{figure}

\section{Simulation of stochastic processes in OpenModelica}

As already mentioned in the introduction, there are no any
pseudorandom numbers generators in OpenModelica. Thus that makes this
system unusable for stochastic processes modeling. However
\texttt{Noise} library
\url{build.openmodelica.org/Documentation/Noise.html} developed by
Klöckner (Klöckner)~\cite{L_Klockner:2014} should be mentioned. The
basis of this library are \texttt{xorshift} generators
(algorithms~\ref{alg:xorshift64star} and~\ref{alg:xorshift128plus}),
written in C. However, an inexperienced user may face a problem,
because one needs compile C-source files first to use that library.

In this article we will describe the procedure required for connection
of external C functions to OpenModelica programme. That will allow the
user to install the Noise library and to connect their own random
number generators. We also provide a minimal working example of
stochastic Wiener process generator and the example of ordinary
differential equation with additive stochastic part.

\subsection{Connection of external C-functions to OpenModelica program}

Let us consider the process of connection of external functions to
modelica program. The relevant section in the official documentation
misses some essential steps that's why it will lead to an error. All
steps we described, had been performed on a computer with Linux Ubuntu
16.04 LTS and OpenModelica 1.11.0-dev-15.

When the code is compiled OpenModelica program is translated to C code
that then is processed by C-compiler. Therefore, OpenModelica has
built-in support of C-functions. In addition to the C language
OpenModelica also supports Fortran (F77 only) and Python
functions. However, both languages are supported indirectly, namely
via wrapping them in the appropriate C-function.

The use of external C-functions may be required for various reasons,
for example implementations of performance requiring components of the
program, using a fullscale imperative programming language, or the use
of existing sourcecode in C.

We give a simple example of calling C-functions from Modelica
program. Let's create two source files: \verb|ExternalFunc1.c| and
\verb|ExternalFunc2.c|. These files will contain simple functions that
we want to use in our Modelica program.

\begin{verbatim}
// File ExternalFunc1.c
double ExternalFunc1_ext(double x)
{
  return x+2.0*x*x;
}
\end{verbatim}
\begin{verbatim}
// File ExternalFunc2.c
double ExternalFunc2(double x)
{
  return (x-1.0)*(x+2.0);;
}
\end{verbatim}
In the directory, where the source code of Modelica program is placed,
we must create two directories: \verb|Resources| and the
\verb|Library|, which will contain \verb|ExternalFunc1.c| and
\verb|ExternalFunc2.c| files. We should then create object files and
place them in the archive, which will an external library. To do this
we use the following command's list:
\begin{verbatim}
gcc -c -o ExternalFunc1.o ExternalFunc1.c
gcc -c -o ExternalFunc2.o ExternalFunc2.c
ar rcs libExternalFunc1.a ExternalFunc1.o
ar rcs libExternalFunc2.a ExternalFunc2.o
\end{verbatim}

To create object files, we use gcc with \verb|-c| option and the
archiver \verb|ar| to place generated object files in the archive. As
a result, we get two of the file \verb|libExternalFunc1.a| and
\verb|libExternalFunc2.a|. There is also the possibility to put all
the needed object files in a single archive.

To call external functions, we must use the keyword
\verb|external|. The name of the wrapper function in Modelica language
can be differ from the name of the external function. In this case, we
must explicitly specify which external functions should be wrapped.
\begin{verbatim}
model ExternalLibraries
        // Function name differs
  function ExternalFunc1
    input Real x;
    output Real y;
                // Explicitly specifying C-function name
                external y=ExternalFunc1_ext(x) annotation(Library="ExternalFunc1");
  end ExternalFunc1;

  function ExternalFunc2
    input Real x;
    output Real y;
                // The functions names are the same
                external "C" annotation(Library="ExternalFunc2");
  end ExternalFunc2;
  
  Real x(start=1.0, fixed=true), y(start=2.0, fixed=true);
equation
  der(x)=-ExternalFunc1(x);
  der(y)=-ExternalFunc2(y);
end ExternalLibraries;
\end{verbatim}

Note that in the annotation the name of the external library is
specified as \verb|ExternalFunc1|, while the file itself is called
\verb|libExternalFunc1.a|. This is not a mistake and the prefix
\verb|lib| must be added to all library's files.

The example shows that the type \verb|Real| corresponds to the C type
\verb|double|. Additionally, the types of \verb|Integer| and
\verb|Boolean| match the C-type \verb|int|. Arrays of type \verb|Real|
and \verb|Integer| transferred in arrays of type \verb|double| and
\verb|int|.

It should be noted that consistently works only call с-functions with
arguments of \verb|int| and \verb|double| types, as well as arrays of
these types. The attempt to use specific c-type, for example,
\verb|long long int| or an unsigned type such as \verb|unsigned int|,
causes the error.

\subsection{Modeling stochastic Wiener process}

Let us describe the implementation of a generator of the normal
distribution and Wiener process. We assume that the generator of
uniformly-distributed random numbers is already implemented in the
functions \verb|urand|. To generate the normal distribution we will
use Box-Muller transformation and Wiener process can be calculated as
cumulative sums of normally-distributed numbers.

The minimum working version of the code is shown below.The key point
is the use of an operator \verb|sample(t_0, h)|, which generates
events using \verb|h| seconds starting from the time \verb|t_0|. For
every event the operator \verb|sample| calls the function \verb|urand|
that returns a new random number.

\begin{verbatim}
  model generator
    Integer x1, x2;
    Port rnd; "Random number generator's port"
    Port normal; "Normal numbers generator's port"
    Port wiener; "Wiener process values port"
    Integer m = 429496729; "Generator modulo"
    Real u1, u2;
  initial equation
    x1 = 114561;
    x2 = 148166;
  algorithm
    when sample(0, 0.1) then
      x1 := urand(x1);
      x2 := urand(x2);
    end when;
    // normalisation of random sequence
    rnd.data[1] := x1 / m;
    rnd.data[2] := x2 / m;
    u1 := rnd.data[1];
    u2 := rnd.data[2];
    // normal generator
    normal.data[1] := sqrt(-2 * log(u1)) * sin(6.28 * u2);
    normal.data[2] := sqrt(-2 * log(u1)) * cos(6.28 * u2);
    // Wiener process
    wiener.data[1] := wiener.data[1] + normal.data[1];
    wiener.data[2] := wiener.data[2] + normal.data[2];
  end generator;
\end{verbatim}

Note also the use of a special variable of type Port which serves to
connect the various models together. In our example we have created
three such variables: \verb|lg|, \verb|normal|, \verb|wiener|. Because
of this, other models can access the result of our generator.

\begin{verbatim}
  connector Port
    Real data[2];
  end Port;
\end{verbatim}

\begin{figure}
  \centering
  \includegraphics[width=0.8\linewidth]{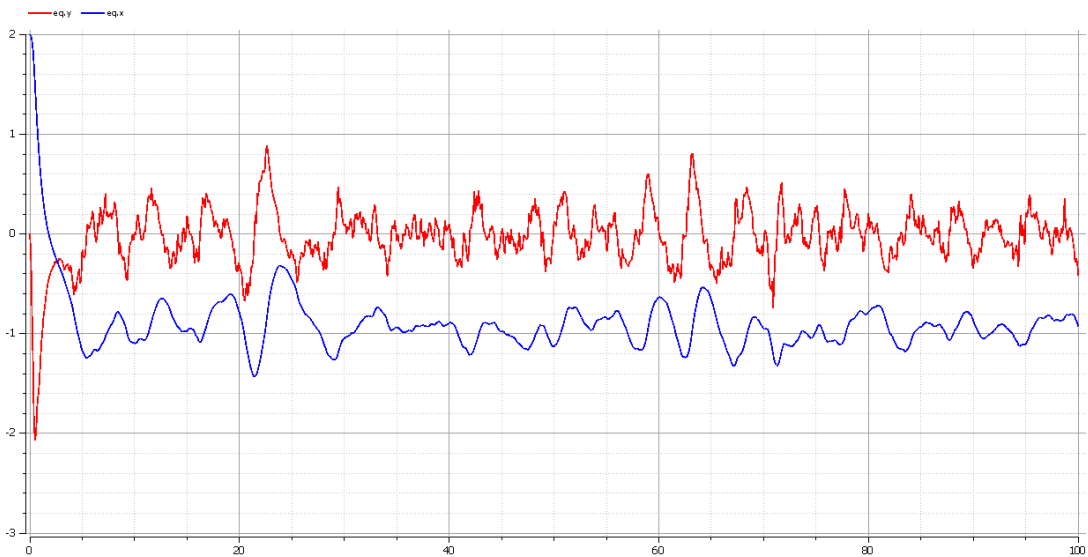}
  \caption{Results of van der Pol–Duffing oscillator simulation. The graphs are created using the functionality OMEditor'a}
  \label{fig:vanderpolduff}
\end{figure}

A minimal working code below illustrates the connection example
between two models. A system of two ordinary differential equations
describes van der Pol–Duffing oscillator with additive stochastic part
in the form of a Wiener process (see~\ref{fig:vanderpolduff}).
\[
  \left\{
  \begin{aligned}
    & \dot{x} = y,\\
    & \dot{y} = x(1.0 - x^{2}) - y + x\cdot W_{t}.
  \end{aligned}
  \right.
\]
It is important to mention that this equation is not
stochastic. Built-in OpenModelica numerical methods do not allow to
solve stochastic differential equations.

\begin{verbatim}
// the model specifies a system of ODE
  model ODE
    Real x, y;
    Port IN;
  initial equation
    x = 2.0;
    y = 0.0;
  equation
    der(x) = y ;
    der(y) = x*(1-x*x) - y + x*IN.data[1];
  end ODE;
  model sim
    generator gen;
    ODE eq;
  equation
    connect(gen.wiener, eq.IN);
  end sim;
\end{verbatim}

\section{Conclusion}

We reviewed the basic algorithms for generating uniformly distributed
pseudo-random numbers. All algorithms were implemented by the authors
in C language and tested using DieHarder utility. The test results
revealed that the most effective algorithms are \texttt{xorshift} and
Mersenne Twister algorithms.

Due to the fact that OpenModelica does not implement bitwise logical
and shifting operators, generators of uniformly distributed
pseudo-random numbers have to be implemented in C language and
connected to the program as external functions. We gave a rather
detailed description of this process, that, as we hope, will fill a
gap in the official documentation.

\begin{acknowledgments}

The work is partially supported by RFBR grants No's 15-07-08795 and 16-07-00556.
Also the publication was supported by the Ministry of Education and
Science of the Russian Federation (the Agreement No~02.A03.21.0008).
The computations were carried out on the Felix computational cluster
(RUDN University, Moscow, Russia) and on the HybriLIT
heterogeneous cluster (Multifunctional center for data storage,
processing, and analysis at the Joint Institute for Nuclear
Research, Dubna, Russia).

\end{acknowledgments}

 \bibliographystyle{elsarticle-num}

\bibliography{bib/cite}

\end{document}

\graphicspath{{image/ru/}}

\floatname{algorithm}{Алгоритм}

\theoremstyle{definition}% Стиль для определений
\newtheorem*{thedefinition*}{Определение}
\newtheorem{thedefinition}{Определение}

\title{Генерация стохастических процессов в OpenModelica}

\author{М. Н. Геворкян}
\email{gevorkyan_mn@rudn.university}
\affiliation{Кафедра прикладной информатики и теории вероятностей,\\
  Российский университет дружбы народов,\\
  ул. Миклухо-Маклая, д. 6, Москва, Россия, 117198}

\author{А. В. Демидова}
\email{demidova_av@rudn.university}
\affiliation{Кафедра прикладной информатики и теории вероятностей,\\
  Российский университет дружбы народов,\\
  ул. Миклухо-Маклая, д. 6, Москва, Россия, 117198}

\author{А. В. Королькова}
\email{korolkova_av@rudn.university}
\affiliation{Кафедра прикладной информатики и теории вероятностей,\\
  Российский университет дружбы народов,\\
  ул. Миклухо-Маклая, д. 6, Москва, Россия, 117198}

\author{Д. С. Кулябов}
\email{kulyabov_ds@rudn.university}
\affiliation{Кафедра прикладной информатики и теории вероятностей,\\
  Российский университет дружбы народов,\\
  ул. Миклухо-Маклая, д. 6, Москва, Россия, 117198}
\affiliation{Лаборатория информационных технологий,\\
  Объединённый институт ядерных исследований,\\
  ул. Жолио-Кюри 6, Дубна, Московская область, Россия, 141980}

\author{Л. А. Севастьянов}
\email{sevastianov_la@rudn.university}
\affiliation{Кафедра прикладной информатики и теории вероятностей,\\
  Российский университет дружбы народов,\\
  ул. Миклухо-Маклая, д. 6, Москва, Россия, 117198}
\affiliation{Лаборатория теоретической физики,\\
  Объединённый институт ядерных исследований,\\
  ул. Жолио-Кюри 6, Дубна, Московская область, Россия, 141980}

\begin{abstract}

\begin{description}
\item[Предпосылки] Язык компонентно-ориентированного моделирования
  \texttt{Modelica} применяется для моделирования сложных процессов,
  задаваемых системой ОДУ. В стандартной библиотеке
  \texttt{OpenModelica} нет средств для генерации псевдослучайных
  чисел, что делает невозможным моделирование стохастических
  процессов.
\item[Цель] Целью данной статьи является краткий обзор некоторого
  числа алгоритмов генерации последовательности равномерно
  распределенных случайных чисел и оценка качества даваемых ими
  последовательностей, а также описание способов реализации некоторых
  из этих алгоритмов в системе OpenModelica.
\item[Методы] Все рассматриваемые алгоритмы реализованы на языке
  \texttt{С}, а результаты их работы протестированы с помощью свободно
  распространяемых тестов \texttt{DieHarder}. Описывается реализация
  алгоритмов на языке \texttt{Modelica} и подключение виде \texttt{C}
  функций.
\item[Результаты] Реализовано и протестировано около девяти
  алгоритмов. С помощью \texttt{DieHarder} выявлены наиболее
  качественные генераторы псевдослучайных чисел. Рассмотрены
  возможности библиотек для \texttt{OpenModelica}: \texttt{Noise} и
  \texttt{AdvancedNoise}.
\item[Выводы] В системе OpenModelica могут быть реализованы
  генераторы равномерно распределенных псевдослучайных чисел, что
  является первым шагом на пути использования OpenModelica для
  моделирования стохастических процессов.
\end{description}

\end{abstract}

  \keywords{Modelica, OpenModelica, генератор случайных величин, винеровский процесс,
    пуассоновский процесс, СДУ}

\maketitle

\section{Введение}

В данной статья рассматривается вопрос генерации равномерно
распределенных псевдослучайных чисел, а также винеровского и
пуассоновского стохастических процессов в среде
OpenModelica~\cite{L_OpenModelica}. OpenModelica является одной из
открытых реализацией языка Modelica~\cite{L_Modelica} (существуют
также другие реализации данного языка~\cite{L_SciLab, L_LMS, L_Dymola,
  L_JModelica, L_WolframSystemModeler, L_MapleSim}). Данный язык
предназначен для моделирования различных систем и процессов, которые
можно представить в виде системы алгебраических или дифференциальных
уравнений. Для численного решения уравнений используется ряд открытых
библиотек~\cite{L_lis:2016, L_LAPACK:2016, L_UMFPACK:2016,
  L_KINSOL:2015}. На данный момент, однако, в стандартной библиотеке
OpenModelica нет средств даже для генерации равномерно распределенных
случайных чисел.

\textbf{В первой части} статьи дается обзор некоторых алгоритмов
генерации псевдослучайных чисел, в том числе описывается
псевдоустройство \texttt{/dev/random} ОС Unix. Для большинства из них
приводится алгоритм в виде псевдокода. Все алгоритмы реализованы
авторами на языке \texttt{C} и частично на языке
\texttt{OpenModelica}, а также протестированы с помощью пакета тестов
\texttt{dieharder}. На основе результатов тестирования выбираются
лучшие алгоритмы для использования.

\textbf{Во второй части} статьи приводятся алгоритмы генерации
нормального и пуассоновско распределений на основе генераторов
равномерно распределенных псевдослучайных чисел. Даются краткие
сведения из теории случайных процессов, аксиоматические определения
пуассоновского и винеровского процессов, а также алгоритмы,
позволяющие генерировать эти процессы на компьютере.

\textbf{Третья часть} статьи носит практическую направленность и
посвящена описанию вызова функций на языке C непосредственно из кода
программы на языке Modelica

\section{Алгоритмы генерации равномерно распределенных псевдослучайных
  чисел}

В данном разделе мы опишем несколько наиболее распространенных
генераторов равномерно распределенных псевдослучайных чисел. Такие
генераторы служат основой для получения последовательностей
псевдослучайных чисел других распределений.

\subsection{Линейный конгруэнтный метод}

Линейный конгруэнтный метод был впервые предложен в 1949 году
Д. Г. Лехмером (D. H. Lehmer)~\cite{L_DKnuth:2004:ru}. Алгоритм~\ref{alg:lcg} задается одной формулой:
\[
x_{n+1} = (a x_{n} + c) \mod m,\;\; n \geqslant 0,
\]
где $m$ --- \emph{модуль} (mask) $m > 1$, $a$ --- \emph{множитель}
(multiplier) $(0 \leqslant a < m)$, $c$ --- \emph{приращение}
$(0 \leqslant c < m)$, $x_0$ --- начальное значение, \emph{зерно}
(seed). Результатом многократного применения данной рекуррентный
формулы является \emph{линейная конгруэнтная последовательность}
$x_{1},\ldots,x_{n}$. Особый случай $c=0$ называется
\emph{мультипликативным} конгруэнтным методом. Для краткого
обозначения данного метода будем использовать аббревиатуру
\texttt{LCG} (\textbf{l}inear \textbf{c}ongruential
\textbf{g}enerator).

\begin{algorithm}[H]
  \caption{\texttt{LCG} линейный конгруэнтный
    генератор}\label{alg:lcg}
  \begin{algorithmic}
    \Require $n$, $seed$ \State $\const\;\;m \leftarrow 2^{64}$ \State
    $\const\;\;a \leftarrow 6364136223846793005$ \State
    $\const\;\;c \leftarrow 1442695040888963407$ \State
    $x_{0} \leftarrow seed$ \For{$i=0$ to $n$} \State
    $x_{i} = (a \cdot x_{i-1} + c) \mod m$
    \EndFor
    \State \Return $\{x_{0}, x_{1}, \ldots, x_{n}\}$
  \end{algorithmic}
\end{algorithm}

Числа $m$, $a$, $c$ называют <<волшебными>> или <<магическими>> так
как их значения задаются в коде программы и выбираются исходя из опыта
применения генератора. Качество генерируемой последовательности
существенно зависит от правильного выбора данных
параметров. Последовательность $\{x\}^{n}_{1}$ периодична и ее период
зависит от числа $m$, которое поэтому должно быть большим. На практике
выбирают $m$ равным машинному слову (для 32-х битной архитектуры —
$2^{32}$ и для 64-х битной --- $2^{64}$). В~\cite{L_DKnuth:2004:ru} рекомендуется выбрать
\[
a=6364136223846793005,\; c=1442695040888963407,\; m=2^{64} =
18446744073709551616.
\]
В статье~\cite{L_Ecuyer:1999} можно найти объемные таблицы с
оптимальными значениями $a$, $b$ и $m$.

Квадратичный конгруэнтный метод
$x_{n} = ( a x_{n-1}^2 + b x_{n-1} + d ) \mod m$ кубический
конгруэнтный метод
$x_{n} = ( a x_{n-1}^3 + b x_{n-1}^2 + c x_{n-1} + d )\mod 2^e$.

В настоящее время линейный конгруэнтный метод представляет по большей
части лишь исторический интерес, так как он генерирует сравнительно
некачественную псевдослучайную последовательность по сравнению с
другими, не менее простыми генераторами.

Авторы реализовали линейный конгруэнтный метод~\ref{alg:lcg} на языке
\texttt{C} и сгенерировали с помощью него последовательность из
$10^{9}$ чисел. Данная последовательность была протестирована с
помощью открытого набора тестов
\texttt{DieHarder}~\cite{L_DieHarder:2013}. В результате генератор
\texttt{LCG} провалил около половины тестов.

\subsection{Метод Фибоначчи с запаздываниями}

Развитием \texttt{LCG} генератора можно считать идею использовать для
генерации $i$-го элемента псевдослучайной последовательности не один,
а несколько предыдущих элементов. Согласно~\cite{L_DKnuth:2004:ru} первый такой генератор был предложен в начале 50-х
годов и основывался на формуле:
\[
        x_{n+1} = (x_{n} + x_{n-1}) \mod m.
\]
Однако на практике он показал себя не лучшим образом. В 1958 году
Дж. Ж. Митчелом (G. J. Mitchell) и Д. Ф. Муром (D. Ph. Moore) был
изобретен намного лучший генератор~\ref{alg:lfg}
\[
        x_{n} = (x_{n-n_{a}} + x_{n - n_{b}}) \mod m,\; n \geqslant \max(n_a, n_b).
\]
Данный генератор получил название генератора Фибоначчи с запаздыванием
(\texttt{LFG}, \textbf{l}agged \textbf{F}ibonacci \textbf{G}enerator).

\begin{algorithm}[H]
  \caption{\texttt{LFG} генератор Фибоначчи с запаздываниями}\label{alg:lfg}
  \begin{algorithmic}
    \State $n_{a} \leftarrow 55$
    \State $n_{b} \leftarrow 24$
    \Require $s_0, s_1,\ldots,s_{n_{b}}$, $n\geqslant  0$
    
    \State $x_{0},x_{1},\ldots,x_{n_{b}}$ $\leftarrow$ $r_0, r_1,\ldots,r_{n_{b}}$
    
    \For{$i = (n_{a}+1)$ to $n$}
      \If{$x_{i-n_a} \geqslant x_{i-n_b}$}
        \State $x_{i} = x_{i-n_a} - x_{i-n_b}$
      \ElsIf{$x_{i-n_a} < x_{i-n_b}$}
        \State $x_i = x_{i-n_a} - x_{i-n_b} + 1$
      \EndIf
    \EndFor
    \State \Return $\{x_{0},x_{1},\ldots,x_{n}\}$
  \end{algorithmic}
\end{algorithm}

Как и в случае \texttt{LCG} генератора, выбор <<магических чисел>>
$n_a$ и $n_b$ сильно влияет на качество генерируемой
последовательности. Авторы предложили использовать следующие
магические числа $n_a$ и $n_b$:
\[
        n_a = 24, n_b = 55.
\]
Д. Кнут~\cite{L_DKnuth:2004:ru, L_DKnuth:1997:en} приводит ряд других
значений, начиная от $(37, 100)$ и заканчивая $(9739, 23209)$ Длина
периода данного генератора в точности равна $2^{e-1}(2^{55} - 1)$ при
выборе $m = 2^e$.

Как видно из алгоритма, для инициализации данного генератора
необходимо использовать не одно начальное значение, а
последовательность из $\max(n_a, n_b)$ случайных чисел.

В открытой библиотеке GNU Scientific Library (GSL)~\cite{L_GSL:2015}
используется \emph{составной мульти-рекурсивный} генератор,
предложенный в статье~\cite{L_Ecuyer:1996}. Данный генератор является
разновидностью \texttt{LFG} и может быть задан следующими формулами:
\[
        \begin{aligned}
                & x_{n} = (a_1 x_{n-1} + a_2 x_{n-2} + a_3 x_{n-3}) \mod m_{1},\\
                & y_{n} = (b_1 y_{n-1} + b_2 y_{n-2} + b_3 y_{n-3}) \mod m_{2},\\
                & z_{n} = (x_{n} - y_{n}) \mod m_{1}.
        \end{aligned}
\]
Составной характер данного алгоритма позволяет получить большой
период, равный $10^{56} \approx 2^{185}$. В GSL используются следующие
значения параметров $a_{i}, b_{i}, m_{1}, m_{2}$:
\[
        \begin{array}{lll}
                a_1 = 0, & b_1 = 86098, & m_1 = 2^{32} - 1 = 2147483647,\\
                a_2 = 63308, & b_2 = 0, & m_2 = 2145483479,\\
                a_3 = -183326, & b_3 = -539608. &
        \end{array}
\]

Еще одни метод, предложный в статье~\cite{L_Ecuyer:1993:1} также
является разновидностью метода Фибоначчи и определяется формулой:
\[
        x_{n} = (a_1 x_{n-1} + a_5 x_{n-5}) \mod 5,
\]
В GSL использованы следующие параметры: $a_1 = 107374182$, $a_2 = 0$,
$a_3 = 0$, $a_4 = 0$, $a_5 = 104480$, $m = 2^{31} - 1 =
2147483647$. Период этого генератора равен $10^{46}$.

Генератор \texttt{LFG}~\ref{alg:lfg} был реализован авторами на языке
\texttt{C} и подвергнут тестированию с помощью
\texttt{DieHarder}~\cite{L_DieHarder:2013}. Генератор показал высокое
качество сгенерированной последовательности ($10^9$ чисел), что дает
основание использовать его при моделировании стохастических процессов.

\subsection{Инверсный конгруэнтный генератор}

Инверсный конгруэнтный метод основан на использовании обратного по
модулю числа.
\[
        x_{i+1} = (a x^{-1}_{i} + b) \mod m
\]
где $a$ --- \emph{множитель} $(0\leqslant a < n)$, $b$ ---
\emph{приращение} $(0\leqslant b < n)$, $x_0$ --- начальное значение
(seed). Кроме того $\text{НОД}(x_0,m)=1$ и $\text{НОД}(a,m)=1$.

Данный генератор превосходит обычный линейный метод, однако сложнее
алгоритмически, так как необходимо искать обратные по модулю целые
числа, что приводит к медленной скорости генерации чисел. Для
вычисления обратного числа обычно применяется расширенный алгоритм
Евклида~\cite[\S 4.3.2]{L_DKnuth:2004:ru}.

\subsection{Генераторы c использованием побитовых операций}

Большинство генераторов, дающих наиболее качественные псевдослучайные
последовательности используют в своих алгоритмах побитовые операции
конъюнкции, дизъюнкции, отрицания, исключающей дизъюнкции
(\texttt{xor}) и побитовые вправо/влево.

\subsubsection{Вихрь Мерсенна}

Считается одним из лучших псевдослучайных генераторов. Разработан в
1997 году Мацумото и Нишимура~\cite{L_Matsumoto:1998:MTE}. Существуют
32-,64-,128-разрядные версии вихря Мерсенна. Свое название алгоритм
получил из-за использования простого числа Мерсенна $2^{19937}-1$. В
зависимости от реализации обеспечивается период вплоть до
$2^{216091}-1$.

Основным недостатком алгоритма является относительная громоздкость и,
как следствие, сравнительно медленная работа. В остальном же данный
генератор обеспечивает качественную псевдослучайную последовательность
и проходит все тесты \verb|DieHarder|. Важным перимуществом является
требование лишь одного инициирующего числа (seed). Вихрь Мерсенна
используется во многих стандартных библиотеках, например в модуле
\texttt{random} языка \texttt{Python 3}~\cite{L_Python3:3.5.1}.

Ввиду громоздкости алгоритма мы не приводим его псевдокод в данной
статье, однако стандартная имплементация алгоритма, созданная Мацумото
и Нишимура свободно доступна по ссылке
\url{http://www.math.sci.hiroshima-u.ac.jp/~m-mat/MT/emt64.html}.

\subsubsection{Генераторы \texttt{XorShift}}

Несколько простых генераторов (алгоритмы~\ref{alg:xorshift64star}
и~\ref{alg:xorshift128plus}, дающих качественную псевдослучайную
последовательность были разработаны в 2003 году Дж. Марсальей
(G. Marsaglia)~\cite{L_xorshift:2003, L_xorshift:2005}.

\begin{minipage}[t]{0.5\textwidth}
        \vspace{-10pt}
        %%%%%%%%%%%%%%%%%%%%%%%%%%%%%%%%%%%%%%%%%%%%%%%%%%%%%%%%%%%%%%%%%%%%%%%%%%%%%
        \begin{algorithm}[H]
                \begin{algorithmic}
                        \caption{Генератор \texttt{xorshift*}}\label{alg:xorshift64star}
                        \Require $n$, $seed$
                        \State $x \leftarrow seed$
                        \State $y_{0} \leftarrow x$
                        \For{$i=1$ to $n$}
                                \State $x$ $\leftarrow$ $x\; \oplus$ $x \gg 12$
                                \State $x$ $\leftarrow$ $x\; \oplus$ $x \ll 25$
                                \State $x$ $\leftarrow$ $x\; \oplus$ $x \gg 27$
                                \State $y_{i} \leftarrow  x \cdot 2685821657736338717$ 
                        \EndFor
                        \State \Return $\{y_{0}, y_{1}, \ldots, y_{n}\}$
                \end{algorithmic}
        \end{algorithm}
        %%%%%%%%%%%%%%%%%%%%%%%%%%%%%%%%%%%%%%%%%%%%%%%%%%%%%%%%%%%%%%%%%%%%%%%%%%%%%
\end{minipage}
\begin{minipage}[t]{0.5\textwidth}
        \vspace{-10pt}
        %%%%%%%%%%%%%%%%%%%%%%%%%%%%%%%%%%%%%%%%%%%%%%%%%%%%%%%%%%%%%%%%%%%%%%%%%%%%%
        \begin{algorithm}[H]
                \caption{Генератор \texttt{xorshift+}}\label{alg:xorshift128plus}
                \begin{algorithmic}
                        \Require $n$, $seed_{1}$, $seed_{2}$
                        \For{$i=1$ to $n$}
                                \State $x \leftarrow seed_{1}$
                                \State $y \leftarrow seed_{2}$
                                \State $seed_{1} \leftarrow y$
                                \State $x = x \oplus (x << 23)$
                                \State $seed_{2} = x \oplus y \oplus (x >> 17) \oplus (y >> 26)$
                                \State $z_{i} \leftarrow seed_{2} + y$
                        \EndFor
                        \State \Return $\{z_{1},\ldots,z_{n}\}$
                \end{algorithmic}
        \end{algorithm}
        %%%%%%%%%%%%%%%%%%%%%%%%%%%%%%%%%%%%%%%%%%%%%%%%%%%%%%%%%%%%%%%%%%%%%%%%%%%%%
\end{minipage}

\subsubsection{Генераторы KISS (Keep It Simple Stupid)}

Еще одно семейство генераторов, дающих не менее качественную
последовательность псевдослучайных чисел~\cite{L_KISS:2011}. Генератор
\texttt{KISS} используется в процедуре \texttt{random\_number()} языка
\texttt{Frotran} (компилятор \texttt{gfortran}~\cite{L_gfortran:2015})

\begin{minipage}[t]{0.5\textwidth}
        \vspace{-10pt}
        %%%%%%%%%%%%%%%%%%%%%%%%%%%%%%%%%%%%%%%%%%%%%%%%%%%%%%%%%%%%%%%%%%%%%%%%%%%%%
        \begin{algorithm}[H]
                \begin{algorithmic}
                        \caption{Генератор \texttt{KISS}}\label{alg:kiss}
                        \Require $n$, $seed_{0},seed_{1},seed_{2},seed_{3}$
                        \State $t$
                        \For{$i=1$ to $n$}
                                \State $seed_{0} \leftarrow 69069 \cdot seed_{0} + 123456$
                                \State $seed_{1} \leftarrow seed_{1} \oplus (seed_{1} << 13)$
                                \State $seed_{1} \leftarrow seed_{1} \oplus (seed_{1} >> 17)$
                                \State $seed_{1} \leftarrow seed_{1} \oplus (seed_{1} << 5)$
                                \State $t \leftarrow 698769069 \cdot seed_{2} + seed_{3}$
                                \State $seed_{3} \leftarrow (t >> 32)$
                                \State $seed_{1} \leftarrow t$
                                \State $x_{i} \leftarrow seed_{0} + seed_{1} + seed_{2}$
                        \EndFor
                        \State \Return $\{x_{1},\ldots,x_{n}\}$
                \end{algorithmic}
        \end{algorithm}
        %%%%%%%%%%%%%%%%%%%%%%%%%%%%%%%%%%%%%%%%%%%%%%%%%%%%%%%%%%%%%%%%%%%%%%%%%%%%%
\end{minipage}
\begin{minipage}[t]{0.5\textwidth}
        \vspace{-10pt}
        %%%%%%%%%%%%%%%%%%%%%%%%%%%%%%%%%%%%%%%%%%%%%%%%%%%%%%%%%%%%%%%%%%%%%%%%%%%%%
        \begin{algorithm}[H]
                \begin{algorithmic}
                        \caption{Генератор \texttt{jKISS}}\label{alg:jkiss}
                        \Require $n$, $seed_{0},seed_{1},seed_{2},seed_{3}$
                        \State $t$
                        \For{$i=1$ to $n$}
                                \State $seed_{0} \leftarrow 314527869 \cdot seed_{0} + 1234567$
                                \State $seed_{1} \leftarrow seed_{1} \oplus (seed_{1} << 5)$
                                \State $seed_{1} \leftarrow seed_{1} \oplus (seed_{1} >> 7)$
                                \State $seed_{1} \leftarrow seed_{1} \oplus (seed_{1} << 22)$
                                \State $t \leftarrow 4294584393\cdot seed_{2} + seed_{3}$
                                \State $seed_{3} \leftarrow (t >> 32)$
                                \State $seed_{1} \leftarrow t$
                                \State $x_{i} \leftarrow seed_{0} + seed_{1} + seed_{2}$
                        \EndFor
                        \State \Return $\{x_{1},\ldots,x_{n}\}$
                \end{algorithmic}
        \end{algorithm}
        %%%%%%%%%%%%%%%%%%%%%%%%%%%%%%%%%%%%%%%%%%%%%%%%%%%%%%%%%%%%%%%%%%%%%%%%%%%%%
\end{minipage}

\subsection{Устройства \texttt{/dev/random} и \texttt{/dev/urandom}}

Перевод комментариев к драйверу
random.c. \url{https://git.kernel.org/cgit/linux/kernel/git/stable/linux-stable.git/tree/drivers/char/random.c?id=refs/tags/v3.15.6#n52}.

Для создания истинно-случайной последовательности чисел с помощью
компьютера, в некоторых Unix системах (в частности GNU/Linux)
используется сбор <<фонового шума>> окружения операционной системы и
аппаратного обеспечения. Источником такого случайного шума являются
моменты времени между нажатия клавиш пользователем (inter-keyboard
timings), различные системные прерывания и другие события, которые
удовлетворяют двум требованиям: не детерминированности и сложной
доступности для измерения внешним наблюдателем.

Неопределённость из таких источников собирается драйвером ядра и
помещается в <<энтропийный пул>>, который дополнительно перемешивается
с помощью алгоритма, похожего на алгоритмы вычисления контрольных
сум. Когда случайные байты запрашиваются системным вызовом, они
извлекаются из энтропийного пула путем взятия SHA хеша от содержимого
пула. Взятие хеша позволяет не показывать внутреннее состояние пула,
так как восстановление содержимого по хешу считается вычислительно
невыполнимой задачей. Дополнительно извлекающая процедура занижает
размер содержимого пула для того, чтобы предотвратить выдачу хеша по
всему содержимому и минимизировать теоретическую возможость
определения его содержимого.

Во вне энтропийный пул доступен в виде символьного псевдоустройства
\texttt{/dev/random}. а также в виде системного вызова:
\begin{lstlisting}[language=C]
  void get_random_bytes(void *buf, int nbytes);
  % \end{minted}
\end{lstlisting}
Устройство \texttt{/dev/random} можно использовать для получения очень
качественной последовательности случайных чисел, однако оно возвращает
число байт, равное размеру накопленного энтропийного пула, поэтому
если требуется неограниченное количесто случайных чисел, то следует
использовать символьное псевдоустройство \texttt{/dev/urandom} у
которого нет данного ограничения, но оно уже генерирует
псевдослучайные числа, качество которых достаточно для большинства не
криптографических задач.

\subsection{Тестирование алгоритмов}
Обзор большого числа критериев оценки качества массива псевдослучайных
чисел можно найти в третей главе книги
Д. Кнута~\cite{L_DKnuth:2004:ru}, а также
статье~\cite{L_Ecuyer:2007} одного из ведущих специалистов по
генераторам псевдослучайных чисел. Все описанные в нашей статье
алгоритмы были реализованы на языке \texttt{C} и протестированы с
помощью набора тестов \texttt{dieharder}, доступного на официальном
сайте автора~\cite{L_DieHarder:2013}. Также этот пакет тестов в ходит
в состав официальных репозиториев многих дистрибутивов GNU/Linux.

\subsubsection{Описание \texttt{dieharder}}
Набор тестов \texttt{dieharder} реализован в виде утилиты командной
строки, которая позволяет тестировать последовательности равномерно
распределенных псевдослучайных чисел. Также \texttt{dieharder} может
использовать любой генератор из библиотеки GSL~\cite{L_GSL:2015} для
генерирования чисел или непосредственного тестирования.
\begin{itemize}
        \item \verb|dieharder -l| --- показать список доступных тестов,
        \item \verb|dieharder -g -1| --- показать список доступных генераторов псевдослучайных чисел, каждому генератору присвоено порядковое число, которое надо указать после флага \verb|-g| для включения нужного генератора.   
        \begin{itemize}
                \item \verb|200 stdin_input_raw| --- считывать стандартный входной бинарный поток,
                \item \verb|201 file_input_raw| --- считывать файл в бинарном формате,
                \item \verb|202 file_input| --- считывать файл в текстовом формате,
                \item \verb|500 /dev/random| --- использовать псевдо-устройство \texttt{/dev/random},
                \item \verb|501 /dev/urandom| --- использовать псевдо-устройство \texttt{/dev/urandom}.
        \end{itemize}
\end{itemize}
Каждое псевдослучайное число должно располагаться на новой строке, а в
первых строках файла необходимо указать: тип чисел (\verb|d| --- целые
числа двойной точности), количество чисел в файле и разрядность чисел
($32$ или $64$ бита). Пример такого файла:
\begin{verbatim}
                type: d
                count: 5
                numbit: 64
                1343742658553450546
                16329942027498366702
                3111285719358198731
                2966160837142136004
                17179712607770735227
\end{verbatim}
Когда такой файл создан можно передать его в \verb|dieharder|
\begin{verbatim}
dieharder -a -g 202 -f file.in > file.out
\end{verbatim}
где флаг \verb|-a| включает все встроенные тесты, а флаг \verb|-f| задает файл для анализа. Результаты тестирования будут сохранены в \verb|file.out|.

\subsubsection{Результаты тестов и выводы}

\begin{center}
    \begin{tabular}{ | l | c | c | c |}
    \hline
    \textbf{Генератор} & \textbf{Провалено} & \textbf{Слабо} & \textbf{Пройдено} \\ \hline\hline
                \texttt{LCG} (линейный конгруэнтный генератор) & 52 & 6 & 55 \\ \hline
                \texttt{LCG2} (составной линейный конгруэнтный генератор) & 51 & 8 & 54 \\ \hline
                \texttt{LFG} (линейный генератор Фибоначчи) & 0 & 2 & 111 \\ \hline
                \texttt{ICG} (инверсный конгруэнтный генератор) & 0 & 6 & 107 \\ \hline
                \texttt{KISS} & 0 & 3 & 110 \\ \hline
                \texttt{jKISS} & 0 & 4 & 109 \\ \hline
                \texttt{XorShift} & 0 & 4 & 109 \\ \hline
                \texttt{XorShift+} & 0 & 2 & 111  \\ \hline
                \texttt{XorShift*} & 0 & 2 & 111 \\ \hline
                \texttt{MT} (вихрь Мерсенна) & 0 & 2 & 111 \\ \hline
                \texttt{dev/urandom} & 0 & 2 & 111              \\ \hline
    \hline
    \end{tabular}
\end{center}

Из генераторов, использующих побитовые операции выделяются генераторы
\texttt{xorshift*}, \texttt{xorshift+} и вихрь Мерсенна. Все они дают
одинаково качественную последовательность. Алгоритм вихря Мерсенна,
однако, намного более громоздок, чем \texttt{xorshift*} или
\texttt{xorshift+}, поэтому для генерирования больших
последовательностей предпочтительней использовать \texttt{xorshift*}
или \texttt{xorshift+}.

Среди генераторов не использующих побитовые операции выделяется
линейный генератор Фибоначчи, который при тестировании дает результаты
на уровне \texttt{XorShift+} и вихря Мерсенна. Однако необходимость
задать минимум 55 начальных значений для инициализации данного
генератора сводит его полезность к минимуму. Инверсный конгруэнтный
генератор показывает немногим худшие результаты, но требует всего одно
число для инициации алгоритма.

\section{Алгоритмы генерации винеровскго и пуассоновского процессов}

Рассмотрим вопрос генерации нормального и пуассоновского
распределений. Выбор именно этих двух распределений мотивирован их
ключевой ролью в теории стохастических дифференциальных
уравнений. Наиболее общий вид таких уравнений использует два случайных
процесса: винеровский и
пуассоновский~\cite{L_Platen_Bruti}. Винеровский процесс позволяет
учесть имплицитную стохастичность моделируемой системы, а
пуассоновский процесс --- внешнее воздействие. Кратко опишем структуру
статьи.

\subsection{Генерирование равномерно распределенных псевдослучайных чисел из единичного промежутка}

Генераторы псевдослучайных равномерно распределенных чисел являются
основой для получения других псевдослучайных
последовательностей. Однако, большинство алгоритмов требуют задания
случайного числа из интервала $[0,1]$, в то время как подавляющее
большинство генераторов равномерно распределенных псевдослучайных
чисел дают последовательность из интервала $[0, m]$, где число $m$
зависит от алгоритма, разрядности операционной системы и процессора.
Чаще всего используют $m \approx 2^{32}$ для 32х битных и
$m \approx 2^{64}$ для 64 битных систем.

Для получения чисел из интервала $[0, 1]$ можно поступить двумя
способами. Во первых можно нормировать пученную последовательность
случайных числе, поделив каждый ее элемент на максимальный элемент
последовательности. Такой подход гарантированно даст $1$ в качестве
одного из случайных чисел. Однако такой способ плох, когда
последовательность псевдослучайных чисел слишком велика и не умещается
в оперативную память. В этом случае лучше использовать второй способ,
а именно поделить каждое сгенерированное число на $m$.

\subsection{Генерирование нормального распределения}

Метод генерации нормально распределенных псевдослучайных величин
предложен в 1958 году Дж. Э. Р. Боксоми
П. Мюллером~\cite{L_BoxMuller:1958} и назван в их честь
\emph{преобразованием Бокса-Мюллера}. Метод основан на простом
порообразовании, задаваемом двумя формулами. Данное преобразование
обычно записывается в двух форматах:
\begin{itemize}
  \item стандартная форма (она как раз и предложена авторами статьи~\cite{L_BoxMuller:1958}),
  \item полярная форма (предложена Дж. Беллом~\cite{L_Bell:1968} и Р. Кнопом~\cite{L_Knop:1969}).
\end{itemize}

\textbf{Стандартная форма.} Пусть $x$ и $y$ --- два независимых, равномерно распределенных псевдослучайных числа из интервала $(0, 1)$, тогда числа $z_1$ и $z_2$ вычисляются по формуле
\[
  z_{1} = \cos(2\pi y)\sqrt{-2\ln{x}},\; z_{2} = \sin(2\pi y)\sqrt{-2\ln{x}}
\]
и являются независимыми псевдослучайными числами, распределенными по
стандартному нормальному закону $\mathcal{N}(0,1)$ с математическим
ожиданием $\mu = 0$ и стандартным среднеквадратичным отклонением
$\sigma = 1$.

\textbf{Полярная форма.} Пусть $x$ и $y$ --- два независимых,
равномерно распределенных псевдослучайных числа из интервала
$[-1, 1]$. Вычислим вспомогательную величину $s = x^2 + y^2$, если
$s>1$ и $s=0$, то значения $x$ и $y$ следует отбросить и проверить
следующую пару. Если же $0 < s \geqslant 1$ тогда числа $z_1$ и $z_2$
вычисляются по формуле
\[
  z_1 = x\sqrt{\dfrac{-2\ln{s}}{s}},\; z_2 = y \sqrt{\dfrac{-2\ln{s}}{s}}
\]
и являются независимыми псевдослучайными числами, распределенными по
стандартному нормальному закону $\mathcal{N}(0,1)$.

При компьютерной реализации данного алгоритма предпочтительней
использовать полярную форму, так как в этом случае приходится
вычислять только одну трансцендентную функцию $\ln$, а не три ($\ln$,
$\sin$ $\cos$), как в стандартном варианте. Это компенсирует даже тот
факт, что часть исходных равномерно распределенных числе отбрасывается
--- полярная версия метода все равно работает быстрее. Пример работы
алгоритма изображен на рисунке~\ref{fig:normal}

Для получения нормального распределения общего вида из стандартного нормального распределения используют формулу $Z = \sigma \cdot z + \mu$, где $z \sim \mathcal{N}(0,1)$, а $Z \sim \mathcal{N}(\mu,\sigma)$.

  % y = - \dfrac{1}{\lambda}\ln(1-x),

\subsection{Генерирование распределения Пуассона}

Для генерирования распределения Пуассона существует большое число
различных алгоритмов~\cite{L_Devroye, L_Ahrens:1974,
  L_Ahrens:1982}. Наиболее простой был предложен
Кнутом~\cite{L_DKnuth:2004:ru}. Для работы
алгоритма~\ref{alg:poisson} необходимо уметь генерировать равномерные
псевдослучайные числа из промежутка $[0,1]$. Пример работы алгоритма
изображе на рисунке~\ref{fig:poisson}
\begin{algorithm}[H]
  \caption{Генератор распределения Пуассона}\label{alg:poisson}
  \begin{algorithmic}
                \Require $seed$, $\lambda$
    
                \State $\Lambda \leftarrow \exp(-\lambda)$, $k \leftarrow 0$, $p \leftarrow 1$, $u \leftarrow seed$
    \Repeat
      \State $k \leftarrow k + 1$
                        \State $u \leftarrow rand(u)$ \Comment{генерируем равномерно распределенное случайное число}
                        \State $p = p \cdot u$
    \Until{$p > \Lambda$}
    \State \Return $k - 1$
  \end{algorithmic}
\end{algorithm}

\begin{figure}
  \centering
  \begin{minipage}[b]{0.48\linewidth}
    \includegraphics[width=1.0\linewidth]{normal.pdf}
    \caption{Нормальное распределение}
    \label{fig:normal}
  \end{minipage}
  \hfill
  \begin{minipage}[b]{0.48\linewidth}
    \includegraphics[width=1.0\linewidth]{poisson.pdf}
    \caption{Распределение Пуассона}
    \label{fig:poisson}
  \end{minipage}
\end{figure}

\subsection{Генерирование пуассоновского и винеровского процессов}

Вышеописанные генераторы нормального и пуассоновского распределений мы
используем для генерации винеровского и пуассоновского стохастических
процессов. Дадим вначале определения данных процессов, а затем
перейдем к описанию алгоритмов.

\subsubsection{Определение винеровского и пуассоновского случайных процессов}

Пусть $(\Omega, \mathscr{A}, P)$ --- вероятностное пространство, где
$\Omega$ --- пространство элементарных событий, $\mathscr{A}$ ---
$\sigma$--алгебра подмножеств $\Omega$ (случайные события), $P$ ---
вероятность или, иначе, вероятностная мера такая что $P(\Omega) = 1$.

\begin{thedefinition*}[см.~\cite{L_Kloeden_Platen, L_Oksendal_en}]
        Семейство случайных величин $X = \{X_{t}, 0 \leqslant t \leqslant T\}$, где $X_{t} \in \mathbb{R}^{d}$ будет называться $d$--мерным стохастическим процессом, где совокупность их конечномерных функций распределения 
        \[
                F_{X_{t_{1}},X_{t_{2}},\ldots,X_{t_{k}}}(x_{i_1},x_{i_2},\ldots,x_{i_k}) = P(X_{t_{1}}\leqslant x_{i_1}, X_{t_{2}} \leqslant x_{i_2}, \ldots, X_{t_{k}} \leqslant x_{i_k})
        \]
        для всех $i_{k} = 1,2,3,\ldots$, $k = 1,2,3,\ldots$, $x_{i} \in \mathbb{R}^{d}$ и $t_{k} \in \mathrm{T}$
\end{thedefinition*}

Пространство состояний $X$ --- $d$--мерное евклидово пространство
$\mathbb{R}^{d}$, $d=1,2,3,\ldots$. Временной интервал $[0, T]$, где
$T>0$. В численных методах рассматривают последовательность моментов
времени $\{t_0,t_1,t_2,\ldots \}$.

\begin{thedefinition*}[см.~\cite{L_Kloeden_Platen, L_Oksendal_en}]
        Случайный кусочно-постоянный процесс $N = \{N_{t}, 0 \leqslant t \leqslant T\}$ с интенсивностью $\lambda > 0$ называется \emph{процессом Пуассона} (пуассоновским) если выполняются следующие свойства.
        \begin{enumerate}
                \item $P\{N_0 = 0\} = 1$, иначе говоря $N_0 = 0$ почти наверное.
                \item $N_t$ --- процесс с независимыми приращениями, то есть $\{\Delta N_{0}, \Delta N_{1}, \ldots\}$ независимые случайные величины; $\Delta N_{t_i} = N_{t_{i+1}} - N_{t_{i}}$ и $0 \leqslant t_{0} < t_{1} < t_{2} < \ldots  < t_{n} \leqslant T$; $\Delta N_{t_i} = N_{t_{i+1}} - N_{t_{i}}$ и $0 \leqslant t_{0} < t_{1} < t_{2} < \ldots  < t_{n} \leqslant T$.
                \item Существует число $\lambda > 0$ такое, что для любого приращения $\Delta N_{i},\; i = 0, \ldots, n-1$, $ E[\Delta N_{i}]= \lambda \Delta t_{i}$.
                \item Если $P(s) = P\{N_{t+s} - N_{t} > 2\}$, то $\lim\limits_{s\to 0} \dfrac{P(s)}{s} = 0$.
        \end{enumerate}
\end{thedefinition*}

\begin{thedefinition*}[см.~\cite{L_Kloeden_Platen, L_Oksendal_en}]
        Случайный процесс $W = \{W_t, 0 \leqslant t \leqslant T\}$ называется скалярным \emph{процессом Винера} (винеровским) если выполняются следующие условия.
        \begin{enumerate}
                \item $P\{W_0 = 0\} = 1$, иначе говоря $W_0 = 0$ почти наверное.
                \item $W_t$ --- процесс с независимыми приращениями, то есть $\{\Delta W_{0}, \Delta W_{1}, \ldots\}$ независимые случайные величины; $\Delta W_{t_i} = W_{t_{i+1}} - W_{t_{i}}$ и $0 \leqslant t_{0} < t_{1} < t_{2} < \ldots  < t_{n} \leqslant T$.
                \item $\Delta W_{i} = W_{t_{i+1}} - W_{t_{i}} \sim \mathcal{N}(0,t_{i+1}-t_{i})$ где $0\leqslant t_{i+1} < t_{i} < T$, $i=0,1,\ldots,n-1$.
        \end{enumerate}
\end{thedefinition*}
Из определения следует, что $\Delta W_{i}$ --- нормально
распределённая случайная величина с математическим ожиданием
$\mathbb{E}[\Delta W_{i}] = \mu = 0$ и дисперсией
$\mathbb{D}[\Delta W_{i}] = \sigma^{2} = \Delta t_{i}$.

Винеровский процесс является моделью \emph{броуновского движения}
(хаотического блуждания). Если рассмотреть процесс $W_t$ в те моменты
времени $0 = t_{0} < t_{1} < t_{2} < \ldots < t_{N-1} < t_{N}$ когда
он испытывает случайные аддитивные изменения, то непосредственно из
определения следует:
$ W_{t_{1}} = W_{t_{0}} + \Delta W_0,\; W_{t_{2}} = W_{t_{1}} + \Delta
W_1,\; \ldots,\; W(t_{N}) = W(t_{N-1}) + \Delta W_{n-1}, $ где
$\Delta W_{i} \sim \mathcal{N}(0,\Delta t_{i})$,
$\forall i = 0,\ldots,n-1$.

Запишем $W_{t_{n}}$ в виде накопленной суммы:
$ W_{t_{n}} = W_{t_{0}} + \sum\limits_{i=0}^{n-1}\Delta W_{i} $ и
учтем, что $\mathbb{E}[\Delta W_{i}] = 0$ и
$\mathbb{D}[\Delta W_{i}] = \Delta t_{i}$ можно показать, что сумма
нормально распределенных случайных чисел $\Delta W_{i}$ также является
нормально распределенным случайным числом:
\[
        \mathbb{E}\sum\limits_{i=0}^{n-1}\Delta W_{i} = 0, \;\; \mathbb{D}\sum\limits_{i=0}^{n-1}\Delta W_{i} = \sum\limits_{i=0}^{n-1} \Delta t_{i} = T,\;\; \sum\limits_{i=0}^{n-1}\Delta W_{i} \sim \mathcal{N}(0,T).
\]

Многомерный винеровский процесс
$\mathbf{W}_{t}\colon \Omega\times[t_{0},T]\to \mathbb{R}^{m}$
определяют как случайный процесс составленный из совместно независимых
одномерных винеровских процессов $W^{1}_t,\ldots,W^{m}_t$. Приращения
$\Delta W^{\alpha}_{i},\;\forall \alpha = 1,\ldots,m$ являются
совместно независимыми нормально распределенными случайными
величинами. С другой стороны вектор $\Delta W^{\alpha}_{i}$ можно
представить как многомерную нормально распределенную случайную
величину с вектором математических ожиданий $\mu = 1$ и диагональной
матрицей ковариаций.

\subsubsection{Генерирование винеровского процесса}

Для симулирования одномерного винеровского процесса необходимо
сгенерировать $N$ нормально распределенных случайных чисел
$\varepsilon_{1}, \ldots, \varepsilon_{N}$ и построить их накопленные
суммы $\varepsilon_{1}$, $\varepsilon_{1} + \varepsilon_{2}$,
$\varepsilon_{1} + \varepsilon_{2} + \varepsilon_{3}$ ... В результате
мы получим \emph{выборочную траекторию} винеровского процесса
$W(t)$~см.~рис.~\ref{fig:wienerprocess}.

В случае многомерного случайного процесса следует сгенерировать уже
$m$ последовательностей из $N$ нормально распределенных случайных
величин.

\subsubsection{Генерирование пуассоновского процесса}

Симулирование пуассоновского процесса во многом аналогично
винеровскому, но теперь необходимо сгенерировать последовательность из
чисел распределенных по пуассоновскому закону и затем вычислить их
накопленную сумму. График процесса изображен на
рис.~\ref{fig:poissonprocess}. Из графика видно, что пуассоновский
процесс представляет собой скачкообразное изменение числа произошедших
с течением времени событий. От интенсивности $\lambda$ зависит среднее
количество событий за отрезок времени.

Из за такого характерного поведения пуассоновский процесс также
называется скачкообразным, а стохастические дифференциальные уравнения
где он участвует в качестве второго ведущего процесса получили
названия уравнений со скачками~\cite{L_Platen_Bruti}

\begin{figure}
  \centering
  \begin{minipage}[b]{0.48\linewidth}
    \includegraphics[width=1.0\linewidth]{wiener_process.pdf}
    \caption{Винеровский процесс}
    \label{fig:wienerprocess}
  \end{minipage}
  \hfill
  \begin{minipage}[b]{0.48\linewidth}
    \includegraphics[width=1.0\linewidth]{poisson_process.pdf}
    \caption{Пуассоновский процесс}
    \label{fig:poissonprocess}
  \end{minipage}
\end{figure}

\section{Моделирование стохастических процессов в OpenModelica}

Как уже отмечалось во введении, в OpenModelica отсутствуют
псевдослучайные генераторы, что делает эту систему непригодной для
моделирования стохастических процессов. Стоит, однако, отметить
библиотеку \texttt{Noise}
\url{build.openmodelica.org/Documentation/Noise.html}, разрабатываемую
Клёкнером (Klöckner)~\cite{L_Klockner:2014} и другими. В основе данной
библиотеки лежит семейство генераторов \texttt{xorshift}
(алгоритмы~\ref{alg:xorshift64star} и~\ref{alg:xorshift128plus} из
первого раздела), написанных на языке C. Однако неподготовленный
пользователь может столкнуться с проблемой при использовании
библиотеки Noise, так как для ее работы необходимо сперва
скомпилировать исходные файлы библиотек на языке C.

В данной части статьи мы опишем процедуру подключения внешних C
функций к OpenModelica, что позволит пользователю использовать как
библиотеку Noise, так и подключить свои генераторы случайных
чисел. Также мы приведем минимальный рабочий пример генерации
стохастического винеровского процесса в среде OpenModelica и решения
обыкновенного дифференциального уравнения с аддитивно добавленным
стохастическим членом.

\subsection{Подключение внешних C функций к OpenModelica}

Перейдем к рассмотрению процесса подключения внешних функций к
программе на OpenModelica. Соответствующий раздел в официальной
документации при описании данной процедуры упускает несколько
существенных деталей и следование его указаниям приводит к ошибке. Все
описанные нами действия проводились на компьютере под управлением
операционной системы Ubunyu Linux версии 16.04 LTS и OpenModelica
версии 1.11.0-dev-15.

При компиляции кода OpenModelica программа транслируется в C-код,
который затем обрабатывается уже C-компилятором. Поэтому поддержка
C-функций реализована в OpenModelica на уровне архитектуры. Кроме
языка C OpenModelica поддерживает также вызов Fortran (только F77) и
Python функций. Однако оба этих языка поддерживаются косвенным путем,
а именно с помощью обертывания их в соответствующие C-функции.

Использование внешних C-функций может понадобится по разным причинам:
повышение быстродействия ресурсоёмких компонент программы,
необходимость использования полноценного императивного языка
программирования, использование уже существующего C-кода т.д.

Приведем простейший пример вызова C функций из программы написанной на
языке Modelica. Создадим два исходных файла \verb|ExternalFunc1.c| и
\verb|ExternalFunc2.c|. В этих файлах будут содержаться простейшие
функции, которые мы хотим использовать в нашей программе на языке
Modelica.

\begin{verbatim}
// Файл ExternalFunc1.c
double ExternalFunc1_ext(double x)
{
  return x+2.0*x*x;
}
\end{verbatim}
\begin{verbatim}
// Файл ExternalFunc2.c
double ExternalFunc2(double x)
{
  return (x-1.0)*(x+2.0);;
}
\end{verbatim}
В директории, где находится исходный код программы на языке Modelica
необходимо создать каталог \verb|Resources| а в нем каталог
\verb|Library|, в котором будут находится оба наших файла
\verb|ExternalFunc1.c| и \verb|ExternalFunc2.c|. После этого следует
создать объектные файлы и поместить их в архив, который будет служить
подключаемой библиотекой. Для этого следует выполнить ряд команд.
\begin{verbatim}
gcc -c -o ExternalFunc1.o ExternalFunc1.c
gcc -c -o ExternalFunc2.o ExternalFunc2.c
ar rcs libExternalFunc1.a ExternalFunc1.o
ar rcs libExternalFunc2.a ExternalFunc2.o
\end{verbatim}

Для создания объектных файлов мы использовали утилиту компиляции gcc с
опцией -c и архиватор ar для помещения созданных объектных файлов в
архив. В результате мы получим два файла \verb|libExternalFunc1.a| и
\verb|libExternalFunc2.a|. Есть также возможность поместить все
необходимые нам объектные файлы в один архив.

Для вызова внешних функций следует использовать ключевое слово
\verb|external|. Название функции-обертки на языке Modelica может как
совпадать, так и отличаться от названия внешней функции. Во втором
случае необходимо явно указать какую внешнюю функции следует обернуть.
\begin{verbatim}
model ExternalLibraries
        // Название функции не совпадает с названием функции
        // на языке C
  function ExternalFunc1
    input Real x;
    output Real y;
                // Явно указываем название С-функции
                external y=ExternalFunc1_ext(x) annotation(Library="ExternalFunc1");
  end ExternalFunc1;

  function ExternalFunc2
    input Real x;
    output Real y;
                // Названия функций совпадают, поэтому явное указание
                // не требуется
                external "C" annotation(Library="ExternalFunc2");
  end ExternalFunc2;
  
  Real x(start=1.0, fixed=true), y(start=2.0, fixed=true);
equation
  der(x)=-ExternalFunc1(x);
  der(y)=-ExternalFunc2(y);
end ExternalLibraries;
\end{verbatim}

Заметьте, что в аннотации название подключаемой библиотеки указанно
как \verb|ExternalFunc1|, в то время как сам файл называется
\verb|libExternalFunc1.a|. Это не является опечаткой и приставку lib
необходимо добавлять ко всем библиотечным файлам.

Из примера видно, что тип \verb|Real| языка Modelica соответствует
типу \verb|double| языка C. Кроме того тип \verb|Integer| и
\verb|Boolean| соответствуют типу \verb|int|. Массивы типа \verb|Real|
и \verb|Integer| также переводятся в массивы типа \verb|double| и
\verb|int|.

Стабильно работает вызов функций с аргументами типа \verb|int| и
\verb|double|, а также массивами этих типов. Попытка же уточнить
используемый тип, например \verb|long long int| или использовать
беззнаковый тип, например \verb|unsigned int| приводит к ошибке.

\subsection{Моделирование стохастического винеровского процесса}

Опишем реализацию генератора нормального распределения и винеровского
процесса. Будем предполагать, что генератор равномерно-распределенных
случайных величин уже реализован в виде функции \verb|urand|. Для
генерирования нормального воспользуемся вышеописанным преобразованием
Бокса-Мюллера, а элементы последовательности, задающей винеровский
процесс вычислим как кумулятивную сумму нормально-распределенных
чисел.

Минимальный рабочий вариант кода приведен ниже.Ключевым моментом
является использование оператора \verb|sample(t_0, h)|, который
генерирует события через \verb|h| секунд, начиная с момента времени
\verb|t_0|. При каждом срабатывании оператора \verb|sample| вызывается
функция \verb|urand|, которая возвращает новое случайное число.

\begin{verbatim}
  model generator
    Integer x1, x2;
    Port lg; "Порт для линейного генератора"
    Port normal; "Порт для нормально распределенных чисел"
    Port wiener; "Порт для винеровского процесса"
    Integer m = 429496729; "Значение модуля (маски), применяемого в генераторе"
    Real u1, u2;
  initial equation
    x1 = 114561;
    x2 = 148166;
  algorithm
    when sample(0, 0.1) then
      x1 := urand(x1);
      x2 := urand(x2);
    end when;
    // если необходимо, то нормируем псевдослучайные числа
    lg.data[1] := x1 / m;
    lg.data[2] := x2 / m;
    u1 := lcg.data[1];
    u2 := lcg.data[2];
    // нормальный генератор
    normal.data[1] := sqrt(-2 * log(u1)) * sin(6.28 * u2);
    normal.data[2] := sqrt(-2 * log(u1)) * cos(6.28 * u2);
    // Винеровский процесс
    wiener.data[1] := wiener.data[1] + normal.data[1];
    wiener.data[2] := wiener.data[2] + normal.data[2];
  end generator;
\end{verbatim}

Отметим также применение специальной переменной типа Port, которая
служит для соединения различных моделей между собой. В нашем примере
мы создали три таких переменных: \verb|lg|, \verb|normal|,
\verb|wiener|. Благодаря этому, другие модели могут получить доступ к
результату работы нашего генератора.

\begin{lstlisting}[language=modelica]
  connector Port
    Real data[2];
  end Port;
  % \end{minted}
\end{lstlisting}

Ниже представлен минимальный код примера иллюстрирующего соединение
моделей между собой. В качестве примера выбрана система из двух
обыкновенных дифференциальных уравнений, описывающая осциллятор
Ван-дер-Поля–Дуффинга и к ней добавлен стохастический вклад в виде
процесса Винера (см.~\ref{fig:vanderpolduff}).
\[
  \left\{
  \begin{aligned}
    & \dot{x} = y,\\
    & \dot{y} = x(1.0 - x^{2}) - y + x\cdot W_{t}.
  \end{aligned}
  \right.
\]
Отметим особо, что это уравнение не является
стохастическим. Встроенные в OpenModelica численные методы не
позволяют решать стохастические дифференциальные уравнения.

\begin{verbatim}
// модель задает систему ОДУ
  model ODE
    Real x, y;
    Port IN;
  initial equation
    x = 2.0;
    y = 0.0;
  equation
    der(x) = y ;
    der(y) = x*(1-x*x) - y + x*IN.data[1];
  end ODE;
// Эту модель надо запускать на симуляцию
// в ней числа из случайного генератора поступают
// в систему ОДУ
  model sim
    generator gen;
    ODE eq;
  equation
    connect(gen.wiener, eq.IN);
  end sim;
\end{verbatim}

\begin{figure}
  \centering
  \includegraphics[width=0.8\linewidth]{oscill.png}
  \caption{Результаты моделирования осциллятора Ван-дер-Поля–Дуффинга с добавленным стохастическим вкладом. Графики созданы с помощью функционала OMEditor'a}
  \label{fig:vanderpolduff}
\end{figure}

\section{Заключение}

Был дан обзор основным алгоритмам генерации равномерно распределенных
псевдослучайных чисел. Все алгоритмы были реализованы авторами на
языке С и протестированы с помощью утилиты DieHarder. Результаты
тестов выявили, что наиболее эффективными алгоритмами являются
алгоритмы \texttt{xorshift} и вихрь Мерсенна.

В связи с тем, что в OpenModelica не реализованы побитовые логические
операции и операции сдвига, генераторы равномерно распределенных
псевдослучайных чисел следует реализовать на языке C и подключить в
виде внешних функций. Мы дали достаточно подробное описание процесса
подключения внешних C-функций, восполняющее пробелы официальной
документации.
        
Был приведён минимальный рабочий пример моделирования винеровского
стохастического процесса, который был использован для моделирования
осциллятор Ван-дер-Поля–Дуффинга с аддитивным стохастическим членом.
Важно отметить, что такое уравнение не является стохастическим. Для
построения моделей на основе стохастических уравнений среду
OpenModelica необходимо дополнить стохастическими численными методами,
для чего по видимому необходимо внести изменения в исходный код
OpenModelica.

\def\acknowledgmentsname{Благодарности}

\begin{acknowledgments}

Работа частично поддержана грантами РФФИ \textnumero~15-07-08795, 16-07-00556.
Также публикация выполнена при финансовой поддержке Минобрнауки России
(Соглашение \textnumero~02.A03.21.0008). 
Расчёты проведены на вычислительном кластере «Felix» РУДН и на
Гетерогенном вычислительном кластере «HybriLIT» Многофункционального
центра хранения, обработки и анализа данных ОИЯИ.

\end{acknowledgments}

  \ifthenelse{\boolean{xetex}}
  {\bibliographystyle{ugost2008l}}
  {\bibliographystyle{gost2008l}}

\bibliography{bib/cite}